\documentclass[pdftex,a4paper,times,twocolumn]{article}
\pdfoutput=1
\usepackage{geometry}
\usepackage{graphicx}
\usepackage{amsmath}
\usepackage[small,bf]{caption}

\usepackage{amssymb}
\usepackage{cite}
\usepackage{abstract}

\title{Characterisation Studies of Silicon Photomultipliers}

\author{Patrick Eckert, Hans-Christian Schultz-Coulon, Wei Shen, Rainer Stamen, Alexander Tadday\thanks{corresponding author, atadday@kip.uni-heidelberg.de} \\
{\it {\small Kirchhoff-Institut f\"ur Physik, Universit\"at Heidelberg, Im Neuenheimer Feld 227, 69120 Heidelberg, Germany}}}
\date{}

\begin{document}

\twocolumn[
\maketitle
\begin{onecolabstract}
This paper describes an experimental setup that has been developed to measure and characterise properties of Silicon Photomultipliers (SiPM). The measured SiPM properties are of general interest for a multitude of potential applications and comprise the Photon Detection Efficiency (PDE), the voltage dependent cross-talk and the after-pulse probabilities. With the described setup the absolute PDE can be determined as a function of wavelength covering a spectral range from $350$ to $\mathrm{1000\, nm}$. In addition, a method is presented which allows to study the pixel uniformity in terms of the spatial variations of sensitivity and gain. The results from various commercially available SiPMs - three HAMAMATSU MPPCs and one SensL SPM - are presented and compared.
\vspace{0.5cm}
\end{onecolabstract}
]
\saythanks
\section{Introduction}
\label{intro}
Silicon Photomultipliers (SiPM) are novel photon detectors whose potential is presently studied by many groups concerning their applicability in many different fields such as high energy physics calorimetry, astrophysics or medical imaging \cite{andreev,astro,pet}. SiPMs consist of a matrix of typically 1000 independent micro-cells (pixels) per $\mathrm{mm^2}$ which are connected in parallel. Each pixel is formed out of a photodiode and a quench resistor in series. The photodiode is operated a few volts above its breakdown voltage such that electrical breakdown occurs if a photoelectron is generated within the active volume. In order to be sensitive to successive photons every avalanche breakdown is interrupted by the built in quench resistor. The charge carrier triggering the avalanche may either be produced by the process of photon absorption or thermal excitation (thermal noise), or it may be released from a defect in the silicon lattice (after-pulse). During electrical breakdown, these primary charge carriers (photoelectrons) are amplified by the avalanche process such that a high gain ($\mathrm{G\approx10^6}$) is achieved, comparable to that of a conventional photomultiplier tube. As the pixel structure can be produced with high uniformity the number of electrons resulting from an avalanche breakdown is approximately constant, no matter which of the pixels fired, yielding an excellent photoelectron resolution of the device.

Relevant SiPM properties are the low operating voltage (usually smaller than $100\,{\rm V}$), their insensitivity to magnetic fields as well as their compact dimensions. Drawbacks arise from the high thermal noise rate (typically from $100\, \mathrm{KHz}$ up to a few MHz at the half photoelectron threshold) and the occurrence of after-pulses and cross-talk. Here, the latter refers to the initiation of an additional breakdown in a secondary pixel through photons generated in the primary avalanche process of the first pixel. An After-pulse is believed to be caused if an electron gets trapped by a defect in the silicon lattice with subsequent release yielding a second, delayed avalanche, if the time between the electron being trapped and released is longer than the typical recovery time of the pixel.

An important parameter of a photodetector is the efficiency at which photons of a given wavelength can be detected. For a SiPM the Photon Detection Efficiency (PDE) can be factorised in three quantities:

\begin{equation}
\mathrm{PDE= \epsilon_{geo} \cdot QE \cdot \epsilon_{trigger}}.
\label{eq:geiger}
\end{equation}

Here $\mathrm{\epsilon_{geo}}$ denotes the geometrical fill-factor, i.e.\ the ratio of the active to the total area of the device; QE refers to the quantum efficiency, i.e.\ the probability that an electron-hole pair is generated; $\mathrm{\epsilon_{trigger}}$ is the combined probability of electrons and holes to initiate electrical breakdown (Geiger breakdown). A review on the properties and recent advances in the field of SiPMs can be found in \cite{renker}.

In order to quantify the relevant properties of SiPMs a test stand has been set up which allows to measure global parameters of SiPMs such as the absolute PDE over a spectral range from $350\,\mathrm{- 1000\,nm}$ as well as the cross-talk and after-pulse probabilities as a function of the applied bias voltage. In addition, the setup allows to measure position dependent properties of the SiPMs by illuminating single pixels with a focussed laser spot; examples are the spatial uniformity in sensitivity, gain and cross-talk probability. Results of these measurements are presented and discussed for four different SiPM devices, three from HAMAMATSU (MPPCs S10362-11-25C/50C/100C) and one from SensL (SPMMICRO1020X13).
\section{PDE Measurement}
The following sections describe the developed laboratory setup, the statistical analysis method used to determine the absolute PDE as well as the method to measure the wavelength dependent relative spectral sensitivity. Corresponding results are discussed at the end of the section.
\label{setup}
\subsection{Absolute PDE Measurement}
\label{absolute}
\begin{figure}[tbp]
	\centering
	\includegraphics[width=1\linewidth]{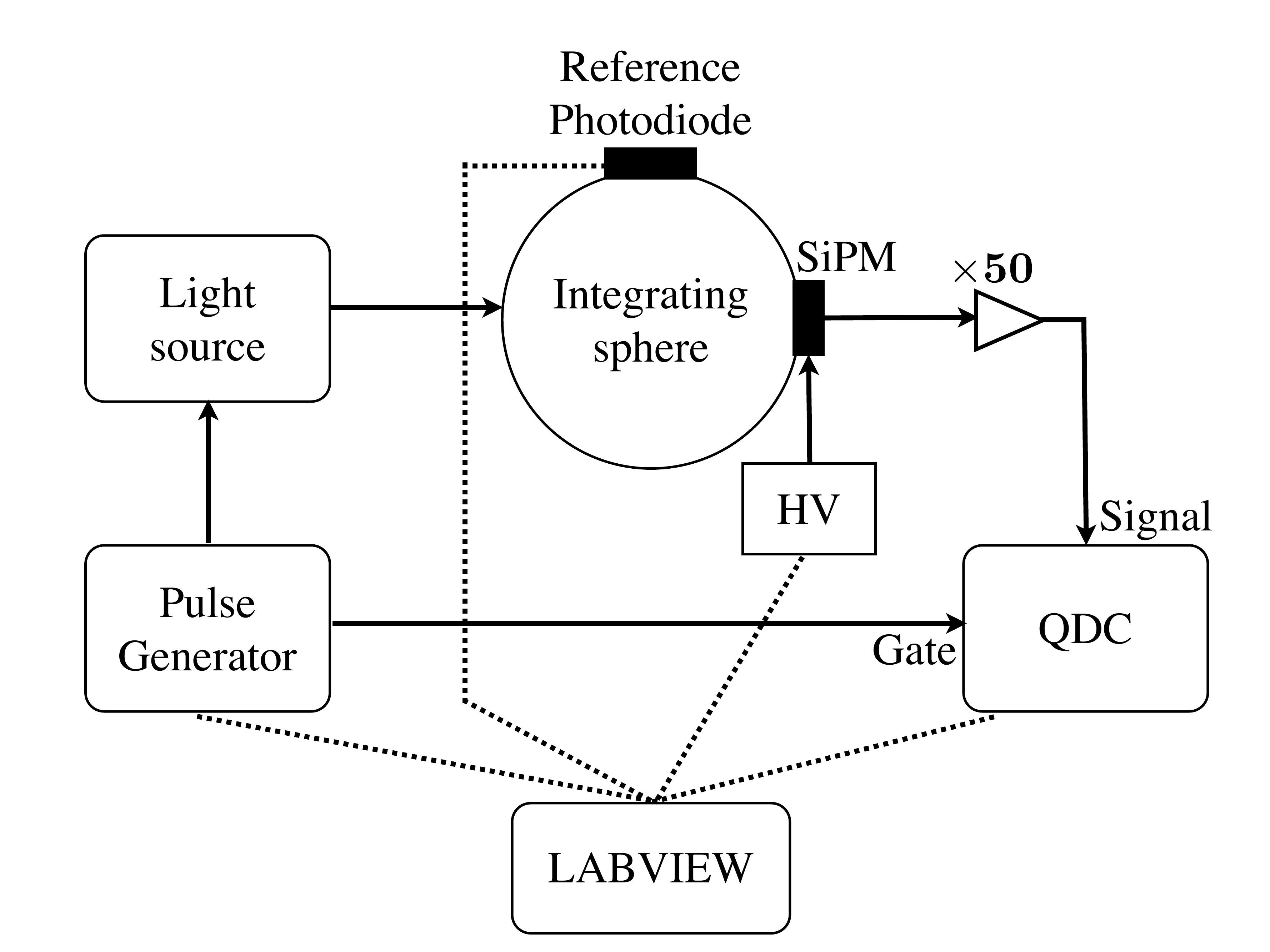}
	\caption{Schematic layout of the PDE measurement setup.}
	\label{fig:setup2}
\end{figure}

\begin{figure}[t]
	\centering
	\includegraphics[width=1\linewidth]{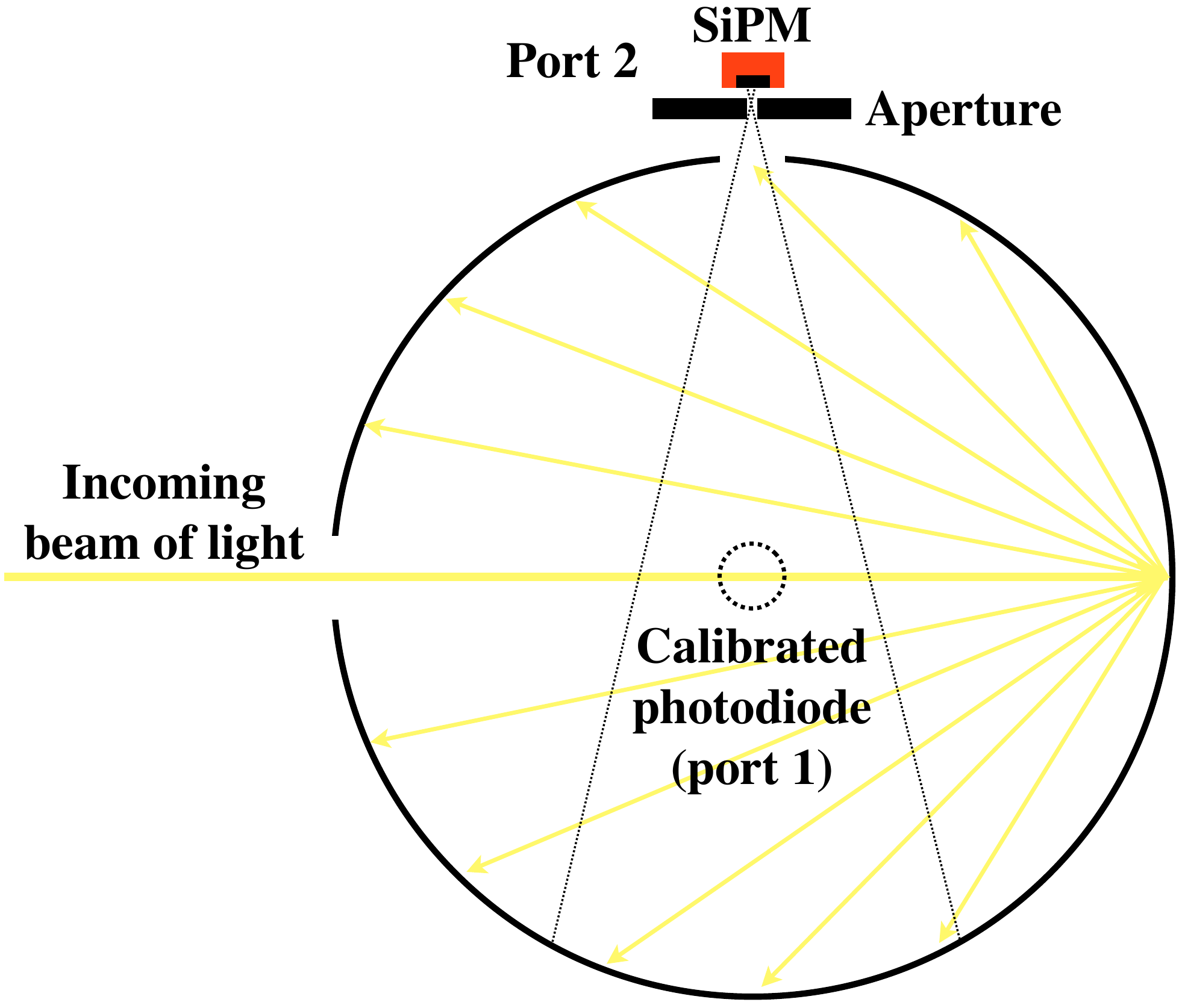}
	\caption{Schematic view of the integrating sphere used in the experimental setup. The connection port of the calibrated photodiode (Port 1) is indicated by the dashed circle. The angle between the individual ports is $90^\circ$, respectively. The dashed lines represent the field of view of the SiPM.}
	\label{fig:sphere}
\end{figure}

\label{sec_exp1}
The layout of the experimental setup is shown in Figure \ref{fig:setup2}; it is controlled by a LABVIEW program to automate the necessary measurement steps. The setup uses an integrating sphere\footnote{Newport Corporation, Model 819D-SL-3.3} (Figure \ref{fig:sphere}) to distribute an incoming beam of light between the SiPM and a NIST\footnote{National Institute of Standards and Technology} certified calibrated photodiode used to determine the absolute amount of light which reaches the SiPM. 
The sphere's cavity has a diameter of $10\,\mathrm{cm}$ and the inner walls are coated with highly diffuse reflective material. This allows to define a reference light source whose output characteristics (light power and angular distribution) do not depend on the direction and intensity profile of the incoming beam of light. Two exit ports are attached to the sphere where the calibrated photodiode (Port 1) and the SiPM (Port 2) can be installed. In order to guarantee that all light enters the $1 \, \mathrm{mm^2}$ active area of the SiPM, it is placed behind an aperture of $0.6\, \mathrm{mm}$ diameter. In order to align the SiPM with respect to the aperture, the device is placed on a movable stage such that its optimal position can be found by moving it in the {\it xy}-plane until the observed signal is maximal.

While the SiPM is placed behind an aperture the calibrated sensor is directly mounted to Port 1 of the sphere; consequently, the amount of light reaching the photodiode is much larger than that entering the SiPM. Hence, a power ratio $R_{0.6}=P_1/P_2$ is defined, which has to be determined experimentally. This is done by comparing the amount of light measured with the calibrated photodiode when it is directly mounted to Port 1, to the situation where it was placed behind the $0.6\, \mathrm{mm}$ aperture of Port 2.  The ratio was measured for different light sources, two laser diodes and two LEDs with different emission wavelengths. The results are shown in Table \ref{table:R}.

The large values measured for $R_{0.6}$ have the advantage that they partially compensate the different sensitivities of the calibrated photodiode and the SiPM, as the SiPM with its much higher gain is capable of detecting single photons, whereas the photodiode ($\mathrm{Gain = 1}$) only generates a measurable output for much higher light intensities.

The emission spectra of the laser-diodes and LEDs in Table \ref{table:R} are determined by placing them in front of a monochromator and measuring the light intensity at the output as a function of the wavelength. The FWHM of these spectra are found to be less than $5\,\mathrm{nm}$ for the laser diodes and between 10 to $20\,\mathrm{nm} $ for the LEDs.
\begin{table}[tb]
\centering 
\begin{tabular}{c c c} 

Light source  & Central wavelength [nm] & Ratio $R_{0.6}$ \\[0.5ex] 

\hline  
LED & 465 & $4200 \pm 20$ \\
Laser diode & 633 & $3852 \pm 18$ \\ 
Laser diode & 775 & $4328 \pm 7$\\
LED & 870 & $4625 \pm 55$  \\[0.5ex] 
\hline 
\end{tabular}
\caption{Measured power ratios for an aperture of $ \varnothing \, 0.6\,$mm at Port 2. The uncertainties were estimated by measuring the power ratio several times for each wavelength and calculating the standard deviation.} 
\label{table:R} 
\end{table}

For the PDE measurements the laser diodes and LEDs are operated in a pulsed mode ($\sim 2\,$ns pulse width) driven by a pulse generator. The SiPM output signal is amplified by a factor of 50 using a fast voltage amplifier\footnote{Phillips Scientific, Model 774}; the charge is measured by a QDC\footnote{LeCroy Model 2249, Charge integrating ADC} with its integration gate set between 50 to $100\,$ns depending on the pulse shape of the tested SiPM device such that coverage of the full output signal is guaranteed.

Similar setups using integrating spheres for determining SiPM PDEs have been previously used \cite{bonanno,otte} . In \cite{bonanno} a pulse counting method is used to determine the PDE. In contrast to the setup in \cite{otte}, the one described here uses a NIST certified calibrated photodiode as reference and allows PDE measurements over a wide spectral range from 350 to $1000\,$nm in $10\,$nm steps (see section \ref{sec:relative_PDE}). In addition, the measurements presented here correctly take the wavelength dependent power ratio (cf.\ Table 1 and Figure 6) into account which improves the accuracy of the results. 

\subsection{Statistical Analysis}
\label{sec:stat}
\begin{figure}[htbp]
	\centering
	\includegraphics[width=1\linewidth]{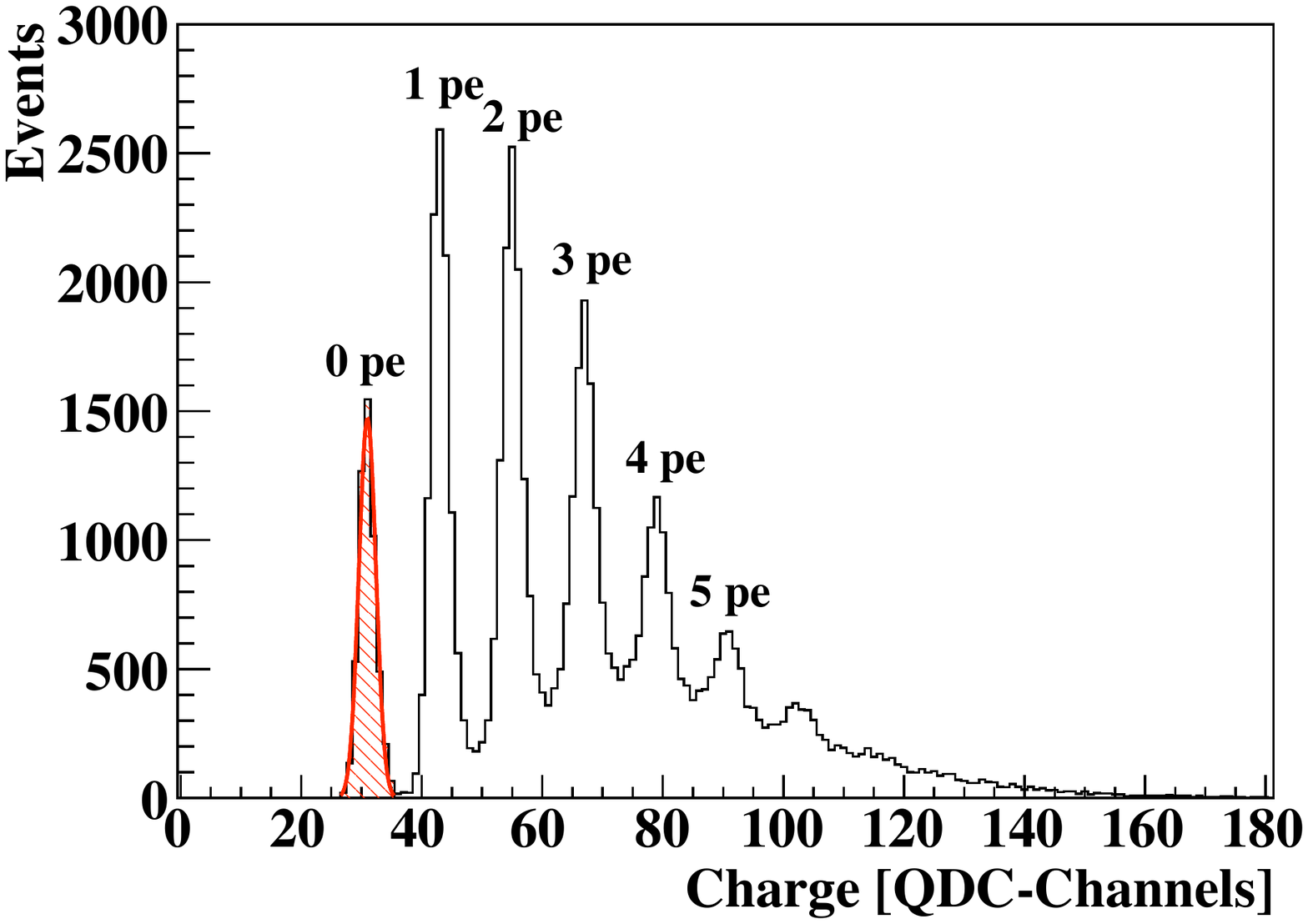}
	\caption{Single Photoelectron spectrum recorded for an MPPC with 1600 pixels. Each peak corresponds to a certain number of photoelectrons (pe).}
	\label{fig:histo}
\end{figure}

The integrated charge values measured by the QDC are filled into a histogram as shown in Figure \ref{fig:histo}. Each of the observed peaks corresponds to a certain number of fired pixels (photoelectrons). The number of photons in a light pulse is expected to be Poisson distributed. The observed photoelectron distribution is distorted due to optical cross-talk and after-pulses.
The area of the first peak (Figure \ref{fig:histo}, shaded region) is, however, unaffected by the cross-talk and after-pulsing. It indicates the number of events, $N_{{\rm ped}}$, in which exactly zero photons have been measured and can be used to determine the PDE without any bias from these two effects. The number of photoelectrons $n_{{\rm pe}}$ can be determined from $N_{\rm{ped}}$ using:

\begin{align}
\label{eq_stat1}
P(0,n_{\rm{pe}}) &= e^{-n_{\rm{pe}}} \nonumber \\
\rightarrow n_{\rm{pe}} &= -ln\left(P(0,n_{\rm{pe}})\right) \nonumber \\
&= -ln\left(\frac{N_{\rm{ped}}}{N_{\rm{tot}} } \right) + ln\left (\frac{N^{\rm{dark}}_{\rm{ped}}}{N^{\rm{dark}}_{\rm{tot}}}\right).
\end{align}

\begin{figure}[htbp]
	\centering
	\includegraphics[width=1\linewidth]{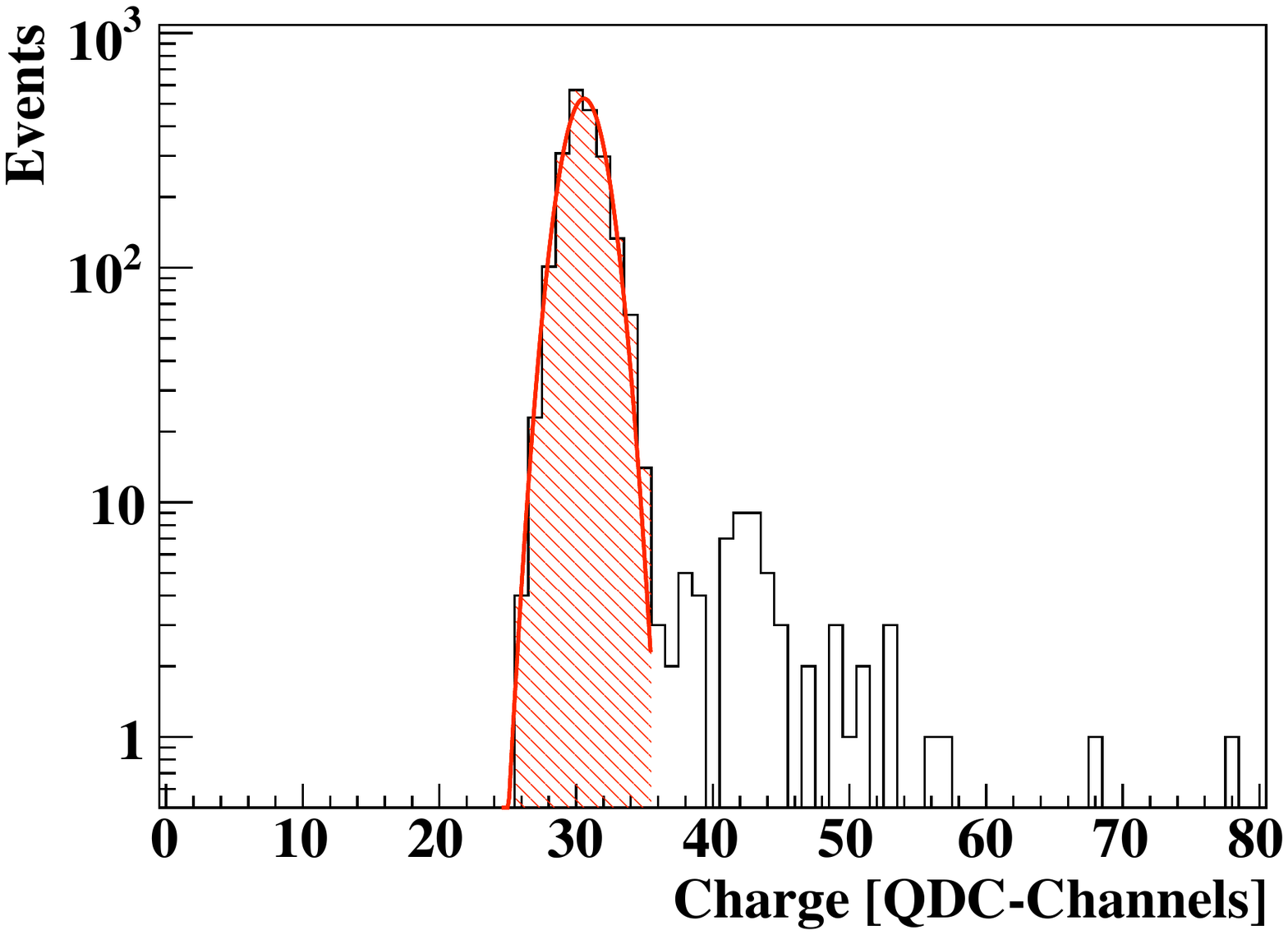}
	\caption{Thermal noise spectrum, recorded for an MPPC with 1600 pixels. The shaded area corresponds to the number of pedestal events.}
	\label{fig:darkrate}
\end{figure}

Here, $P(0,n_{\rm{pe}})$ is the probability to detect zero photons given by a Poisson distribution with mean $n_{\rm{pe}}$, and $N_{\rm{tot}}$ represents the total number of events in the spectrum. $N_{\rm{ped}}$ is determined by fitting a Gaussian function to the pedestal peak and integrating it in a $\pm 3\,\sigma$ interval around the mean value (cf.\ Figure \ref{fig:histo}). The second term in equation \ref{eq_stat1} accounts for the number of detected photoelectrons due to the thermal noise rate. In order to measure $N^{\rm{dark}}_{\rm{ped}}$, which is represented by the shaded area in Figure \ref{fig:darkrate}, the light source was switched off and the QDC readout was triggered arbitrarily.\newline
\begin{figure}[htbp]
	\centering
	\includegraphics[width=1\linewidth]{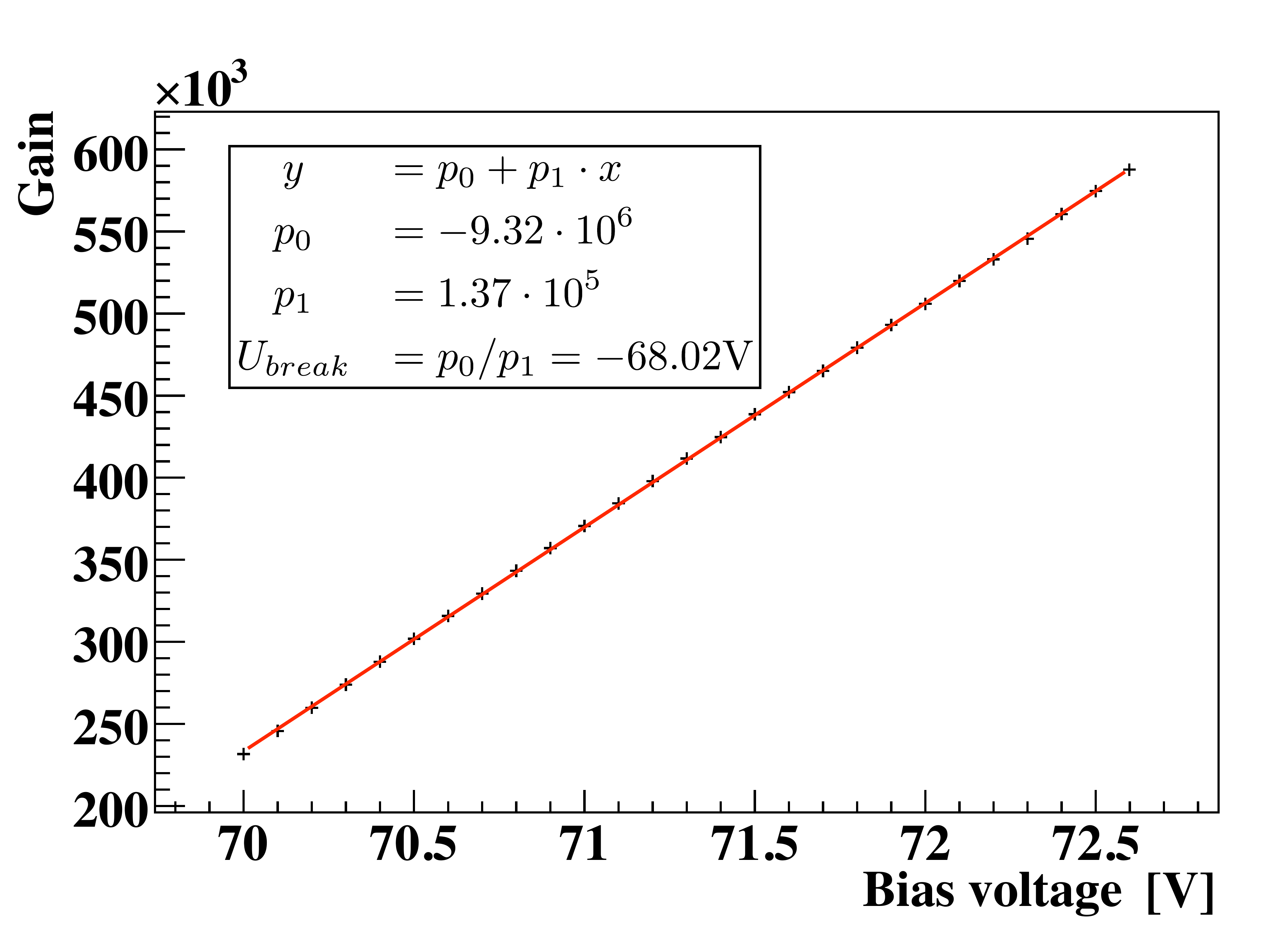}
	\caption{Gain curve for an MPPC with 1600 pixels. The breakdown voltage can be calculated by extrapolating the curve to zero gain.}
	\label{fig:gain}
\end{figure}
The PDE is then calculated from $n_{\rm{pe}}$ considering the power ratio $R_{0.6}$, the period of the light pulses, $T=30 \, \mu s$, and the optical power , $P_{\rm{opt}}$, measured with the calibrated photodiode:

\begin{equation}
{\rm PDE}=\frac{n_{{\rm pe}}\cdot R_{0.6}/T}{P_{{\rm opt}}/(h\cdot \nu)}.
\end{equation}

Where $h$ is Planck's constant and $\nu$ is the frequency of the incoming light.
 
The PDE value varies with the SiPM bias voltage, $U_{\rm{bias}}$. In order to quote the Photon Detection Efficiency as a function of the over voltage, $U_{\rm{over}}=U_{\rm{bias}}-U_{\rm{break}}$, the break down voltage, $U_{\rm{break}}$, has to be determined. This is done by measuring the SiPM gain as a function of $U_{\rm{bias}}$ using the same photon spectra (Fig.\ \ref{fig:histo}) as for the PDE measurements. Hence, photon spectra are recorded for a defined range of bias voltage, and for each bias voltage setting the gain is extracted from the distance of the photo-peaks in the corresponding spectrum. By extrapolating to zero gain using a linear fit (Fig.\ \ref{fig:gain}) the breakdown voltage is determined.

Quoting the PDE value as a function of the over voltage has the advantage that the temperature dependence of the breakdown voltage is taken into account such that a comparison of measurements at different temperatures is possible. This is important since the temperature is only monitored during the measurements but not actively controlled.

\subsection{Relative Sensitivity Measurement}
\label{sec:relative_PDE}
As the statistical analysis requires a pulsed light source it is done only for the wavelengths summarised in Table \ref{table:R}. Thus, in order to measure and characterise the PDE over a wide spectral range a second method is used to determine the relative spectral sensitivity in the range from $\mathrm{350}$ to $\mathrm{1000\,nm}$. The experimental setup is similar to the one shown in Figure \ref{fig:setup2}. As light source a Xenon lamp in combination with a monochromator is used. The SiPM output current is measured with a pico ampere meter\footnote{Keithley Model 6487}.

With this method optical cross-talk and after-pulses cannot be separated, and the measured values of the spectral sensitivity are higher than the actual PDE. Therefore, they have to be normalised to a reference measurement and only the relative shape of the measured curves is used.

For the measurement of the relative sensitivity it is not important that the total amount of light exiting the integrating sphere hits the SiPM. To make use of the full dynamical range by illuminating the entire active area, a slightly wider aperture with a diameter of $0.8\,\mathrm{mm}$ is used. Correspondingly, the power ratio $R_{0.8}$ (defined in the same way as $R_{0.6}$ in section \ref{absolute}) needs to be determined in order to correct for the different light intensities entering the SiPM at Port 1 and the reference photodiode at \mbox{Port 2}.

\begin{figure}[tbp]
	\centering
	\includegraphics[width=1\linewidth]{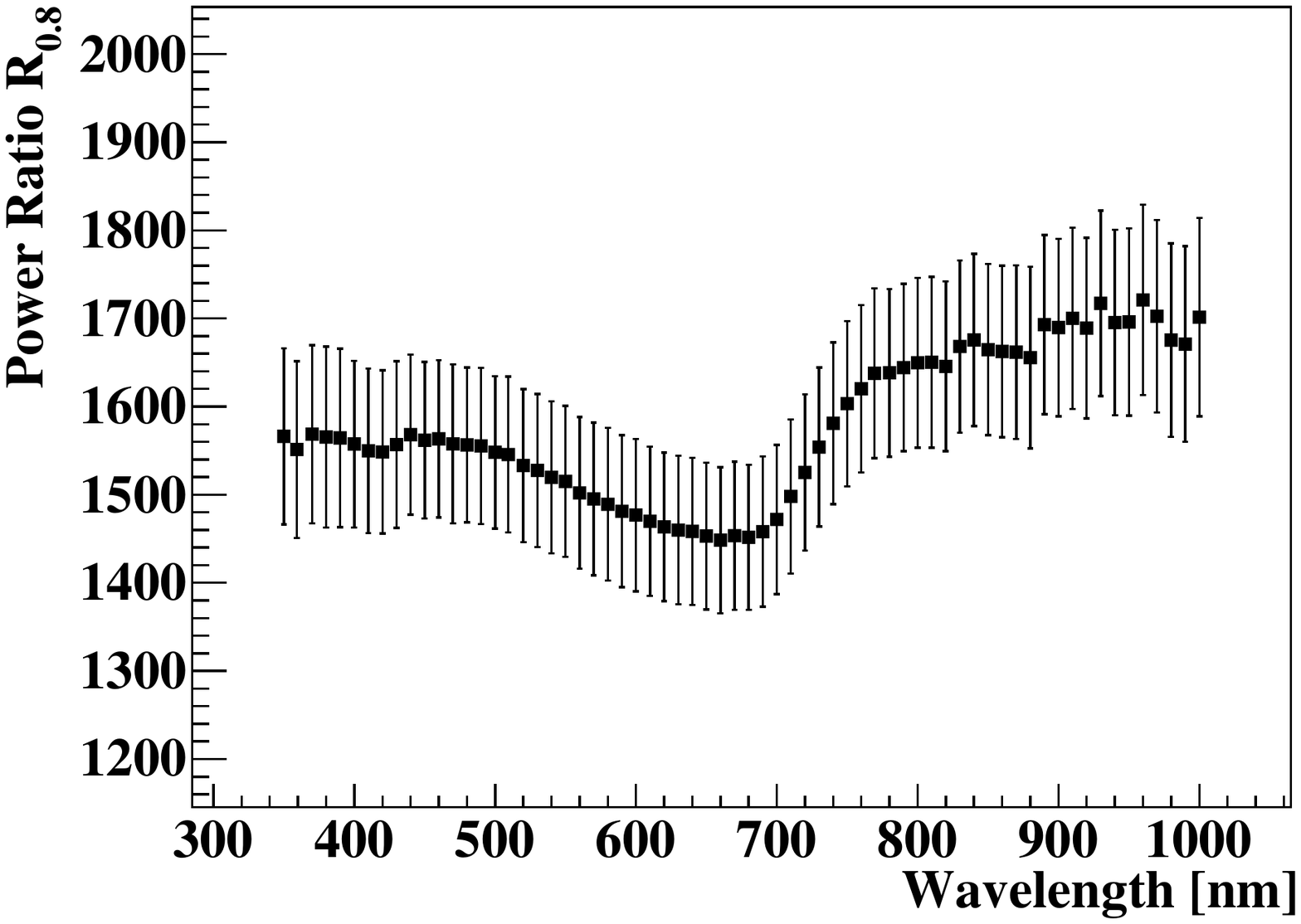}
	\caption{Power ratio $R_{0.8}$ for an aperture of diameter $ \varnothing 0.8\, {\rm mm} $ at Port 2 as a function of the wavelength.}
	\label{fig:ratio}
\end{figure}

As $R_{0.6}$ is observed to be wavelength dependent (cf.\ Tab.\ \ref{table:R}), the power ratio $R_{0.8}$ has to be measured as a function of wavelength as well. The measurement is done by placing a calibrated sensor at each of the two ports of the integrating sphere; $R_{0.8}$ is then given by the ratio of the detected light outputs. The errors are estimated taking the calibration uncertainties and measurement fluctuations into account; for the latter the measurements are repeated several times. The results are shown in Figure  \ref{fig:ratio}. As for $R_{0.6}$ (cf.\ Table \ref{table:R}) the variation of the power ratio is in the order of $10\,\%$ caused by the wavelength dependent reflection properties of the metal surface of the aperture.

In order to guarantee a linear response of the SiPM the light intensities are kept low avoiding the non-linear range of these sensors. The relative spectral sensitivity was calculated using:

 \begin{equation}
 S=\frac{I_{\rm{SiPM}}\cdot R_{0.8} /(q_{\rm{e}}\cdot G)}{P_{\rm{opt}}/(h\cdot  \nu)},
 \end{equation}
 
where  $I_{\rm{SiPM}}$ represents the measured SiPM output current, $G$ the gain of the SiPM and $q_{\rm{e}}$ the elementary charge.
\subsection{Results of the PDE Measurement}
\begin{figure*}[htbp]
	\centering
	\includegraphics[width=0.49\linewidth]{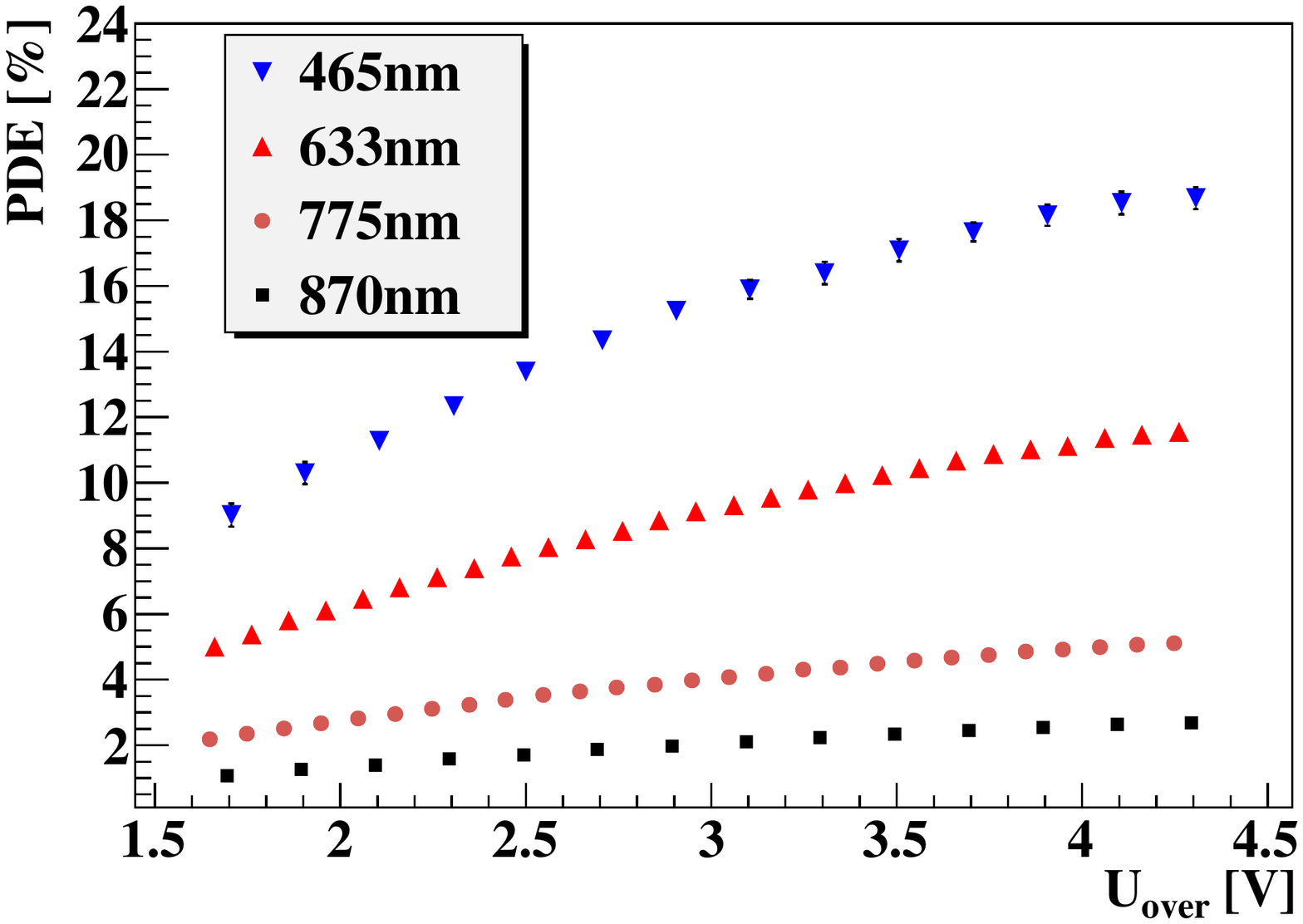}
	\includegraphics[width=0.49\linewidth]{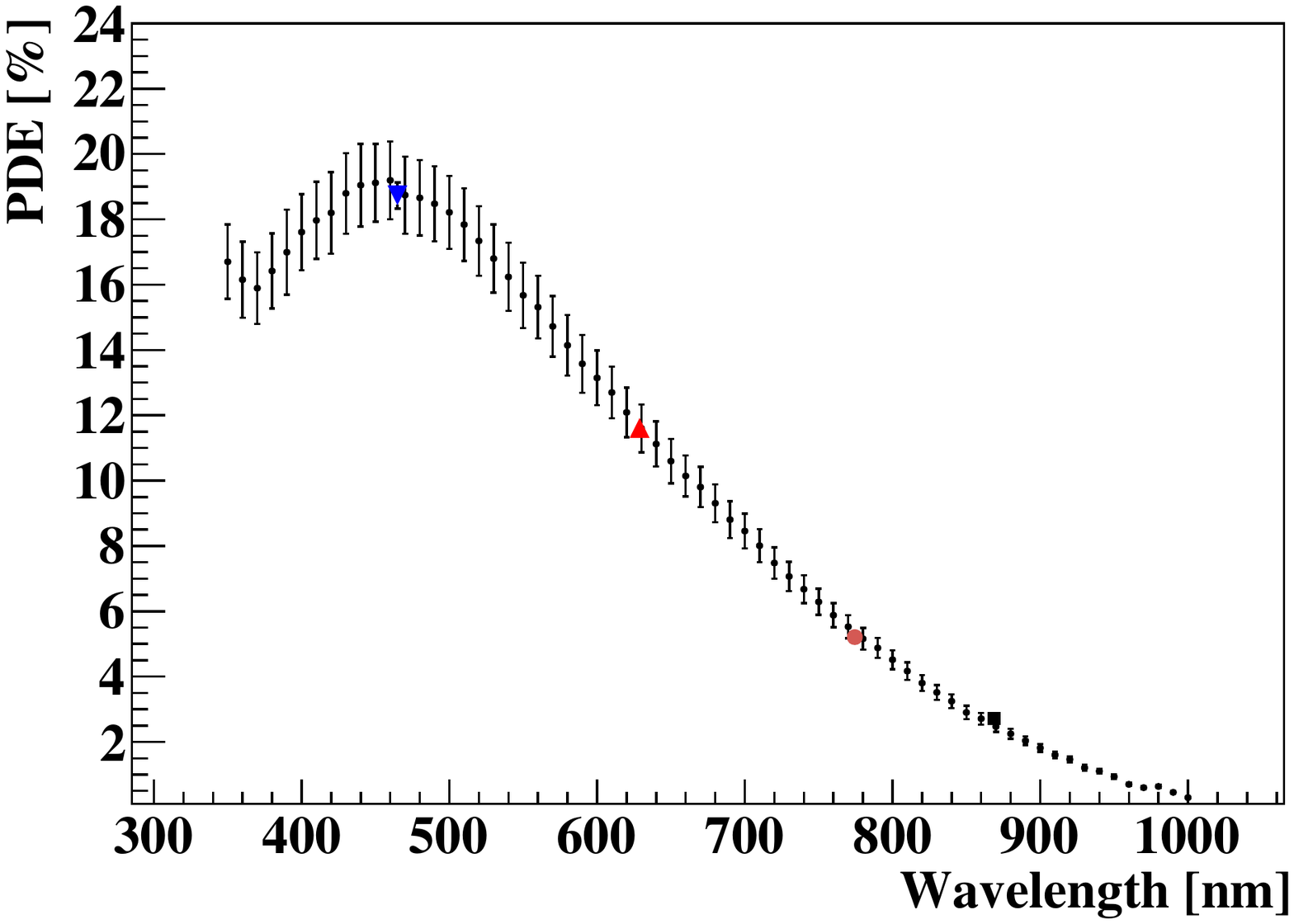}
	\caption{{\bf (left)} Photon detection efficiency of the HAMAMATSU S10362-11-025C as a function of the over voltage for different wavelength of light. {\bf (right)} PDE as a function of the wavelength at an over voltage of $\mathrm{U_{\rm{over}}=(4.3 \pm 0.05)\,V}$ at room temperature ($25\pm1.5 \, ^{\circ}\mathrm{C}$).}
	\label{fig:pde1600}

	\includegraphics[width=0.49\linewidth]{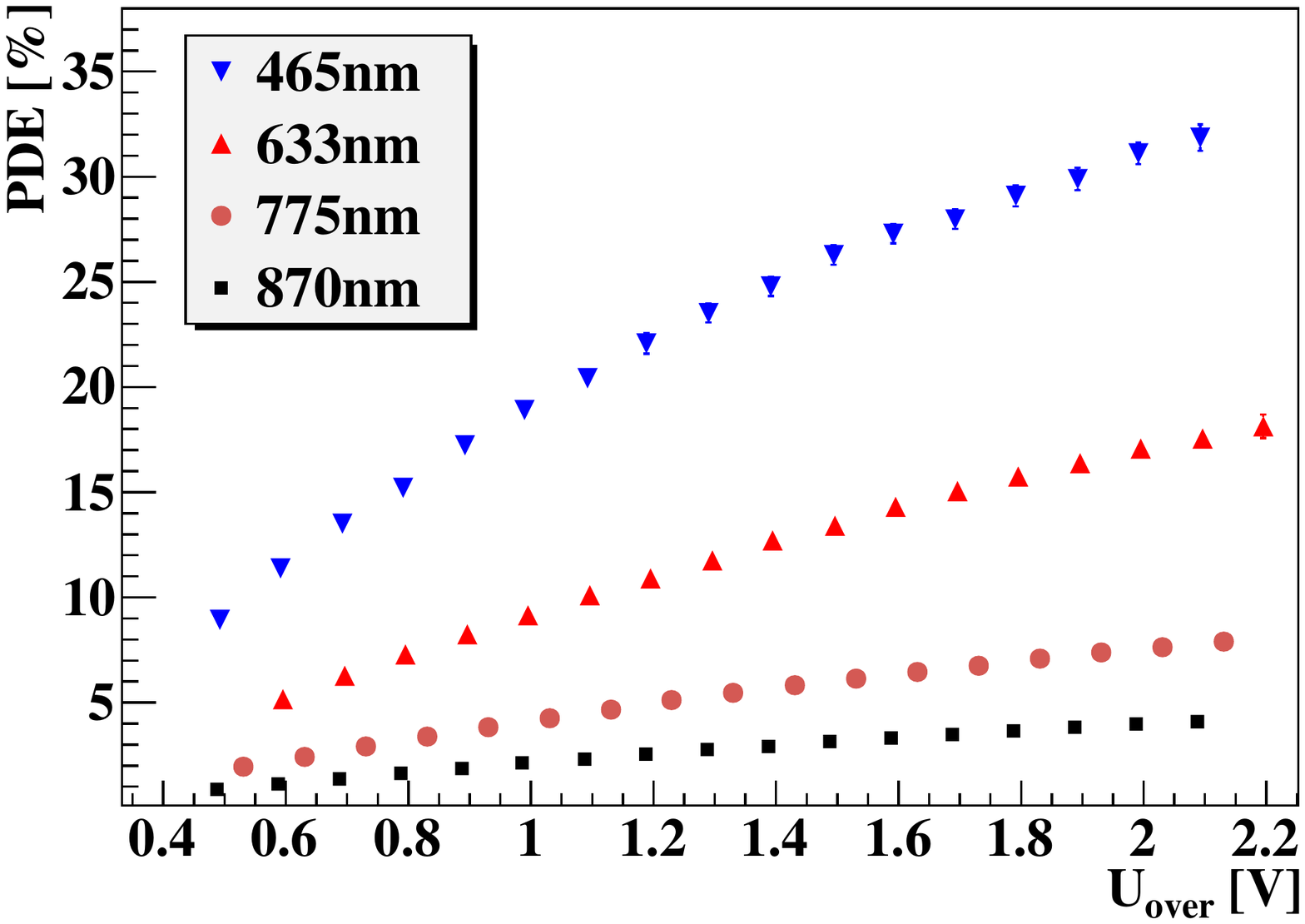}
	\includegraphics[width=0.49\linewidth]{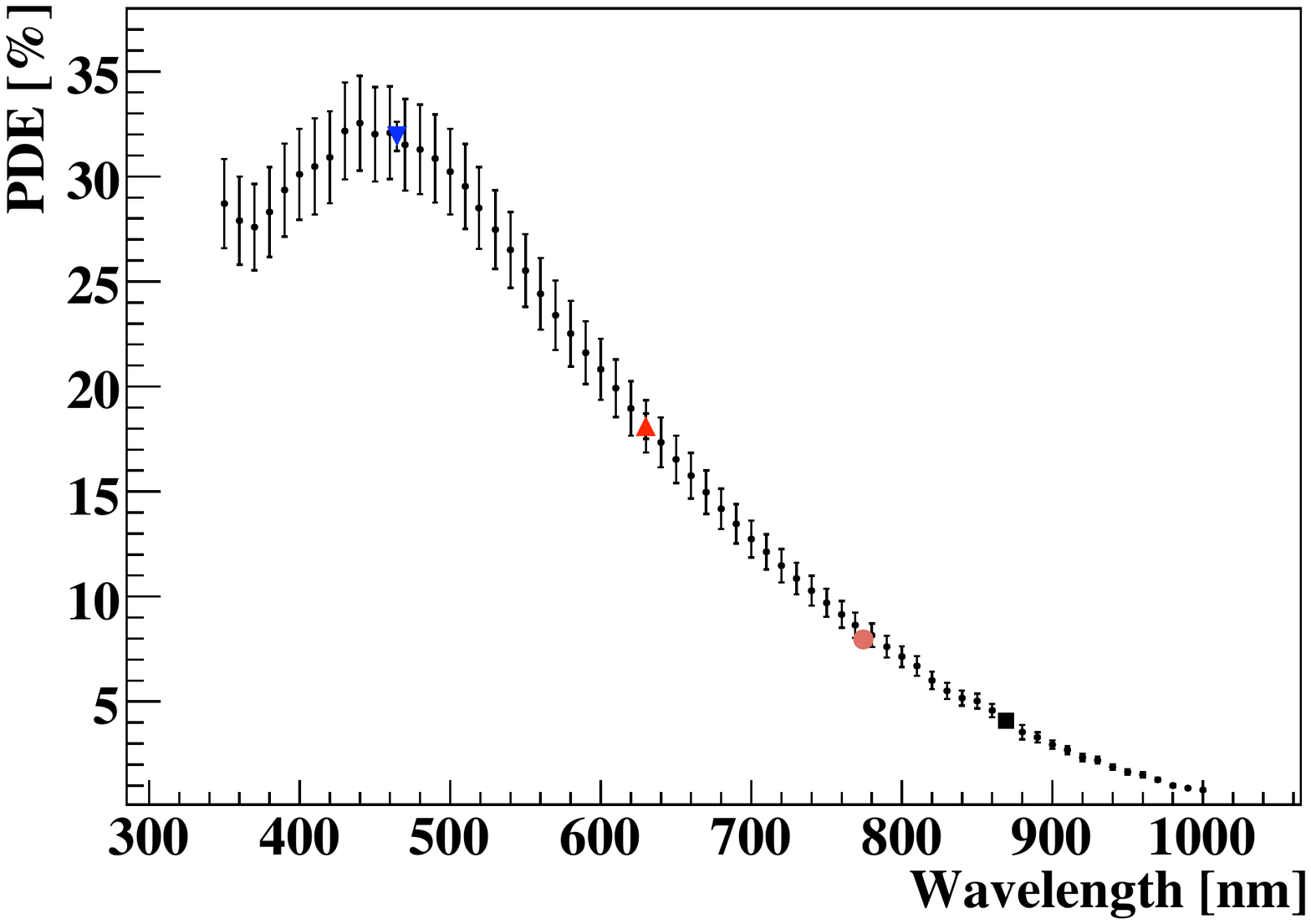}
	\caption{{\bf (left)} Photon detection efficiency of the HAMAMATSU S10362-11-050C as a function of the over voltage for different wavelength of light. {\bf (right)} PDE as a function of the wavelength at an over voltage of $\mathrm{U_{\rm{over}}=(2.15 \pm 0.05)\,V}$ at room temperature ($25\pm1.5 \, ^{\circ}\mathrm{C}$).}
	\label{fig:pde400}

	\includegraphics[width=0.49\linewidth]{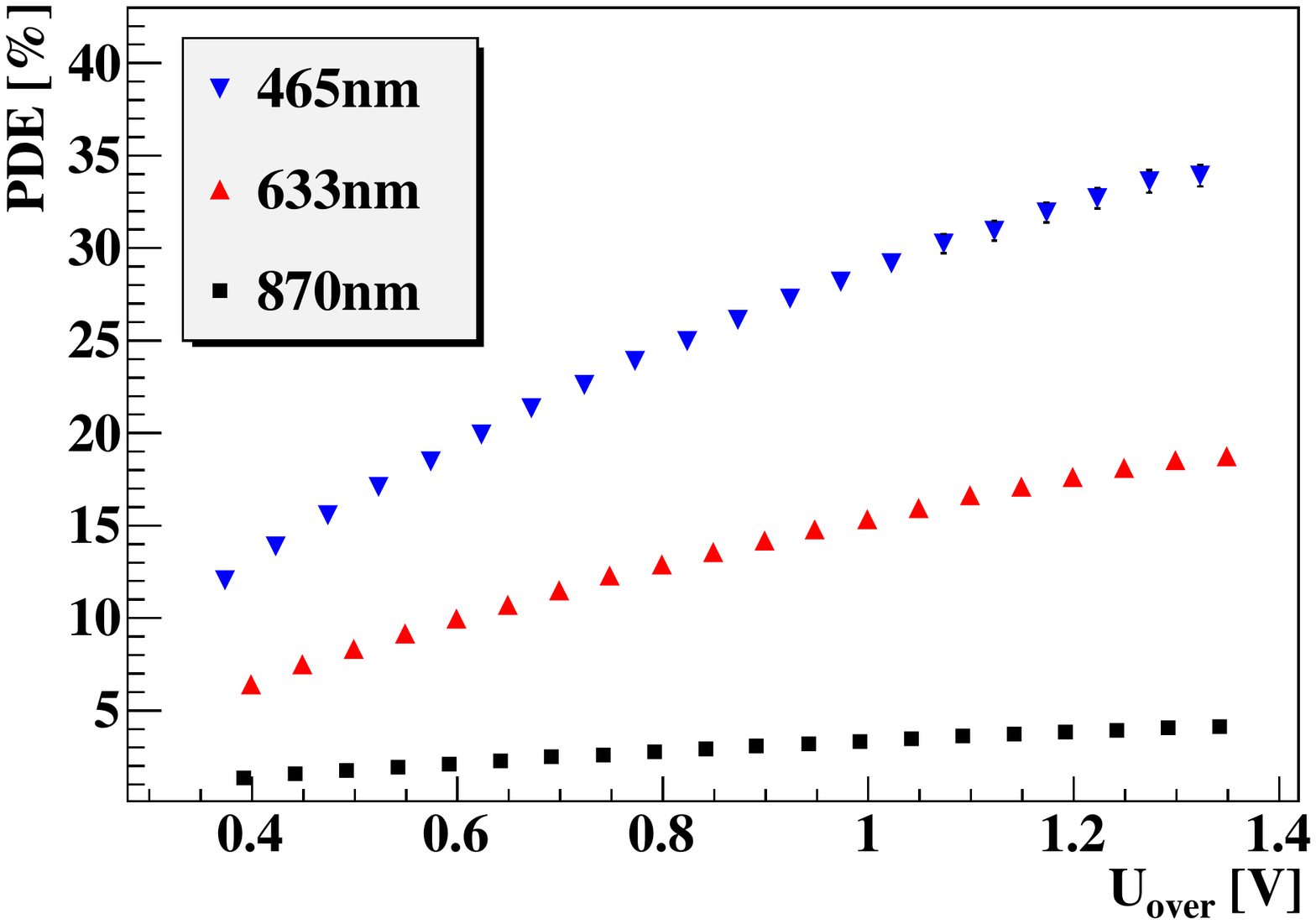}
	\includegraphics[width=0.49\linewidth]{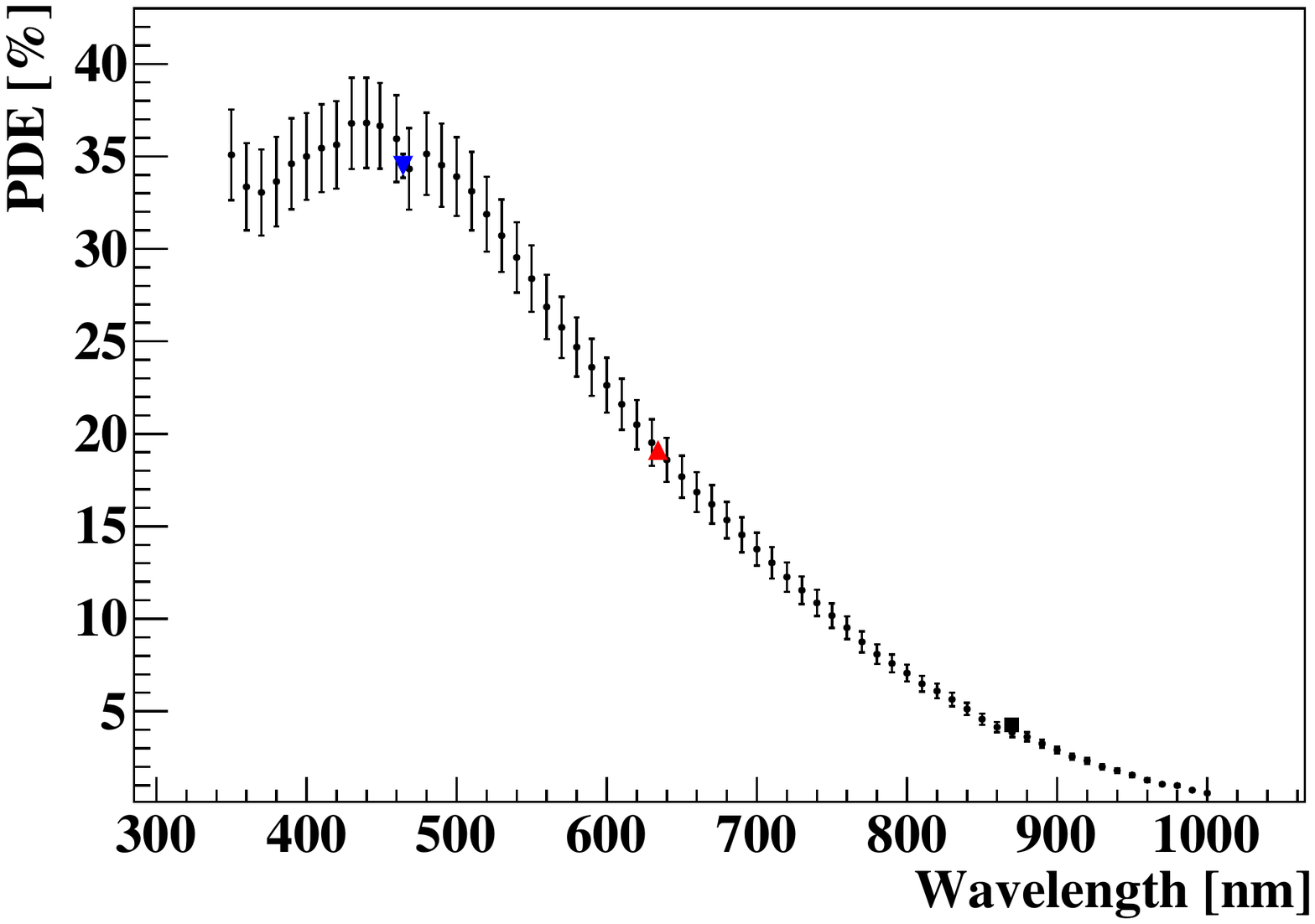}
	\caption{{\bf (left)} Photon detection efficiency of the HAMAMATSU S10362-11-100C as a function of the over voltage for different wavelength of light. {\bf (right)} PDE as a function of the wavelength at an over voltage of $\mathrm{U_{\rm{over}}=(1.3 \pm 0.05)\,V}$ at room temperature ($25\pm1.5 \, ^{\circ}\mathrm{C}$).}
	\label{fig:pde100}
\end{figure*}

\begin{figure*}[htbp]	
	\includegraphics[width=0.49\linewidth]{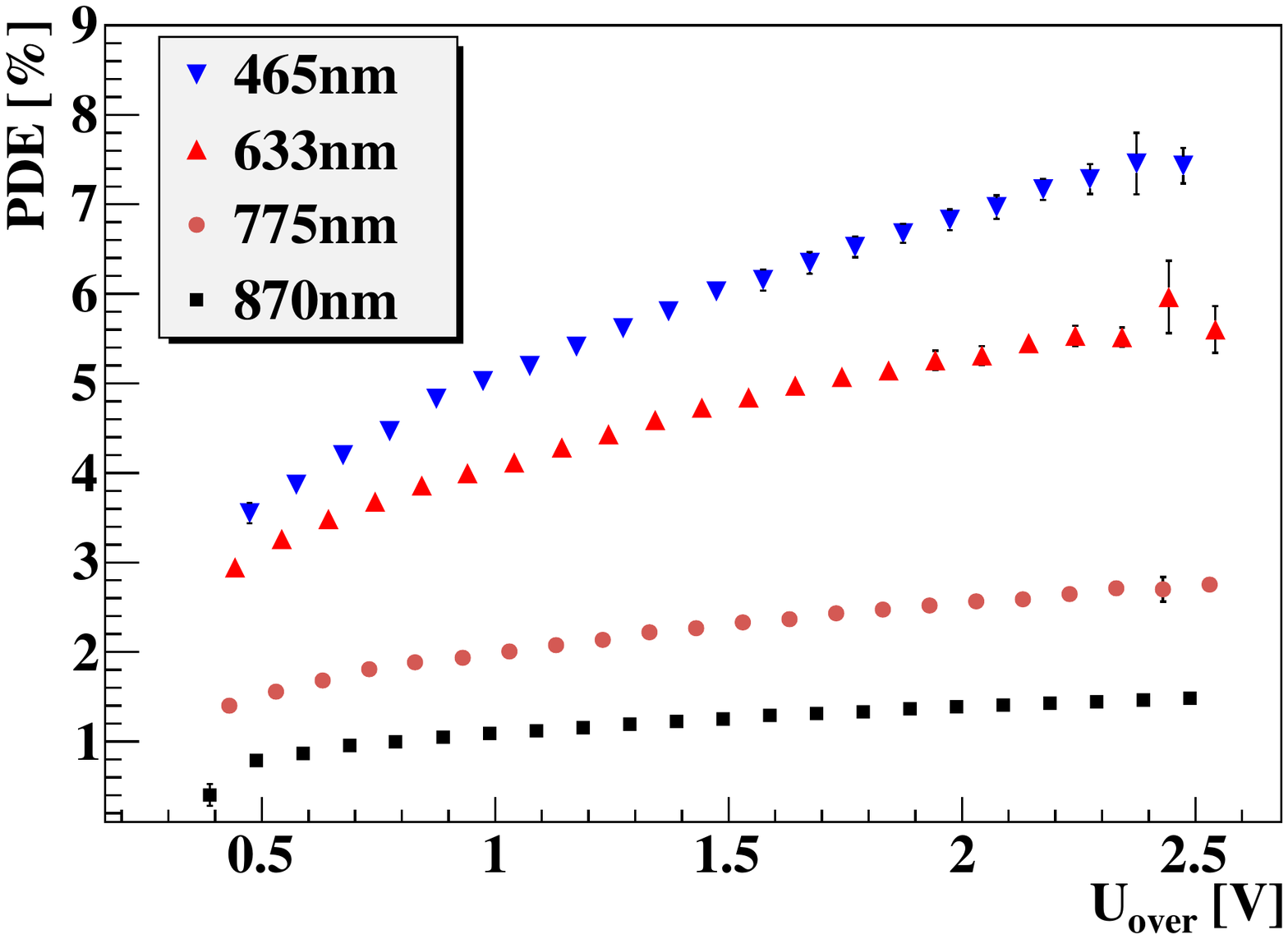}
	\includegraphics[width=0.49\linewidth]{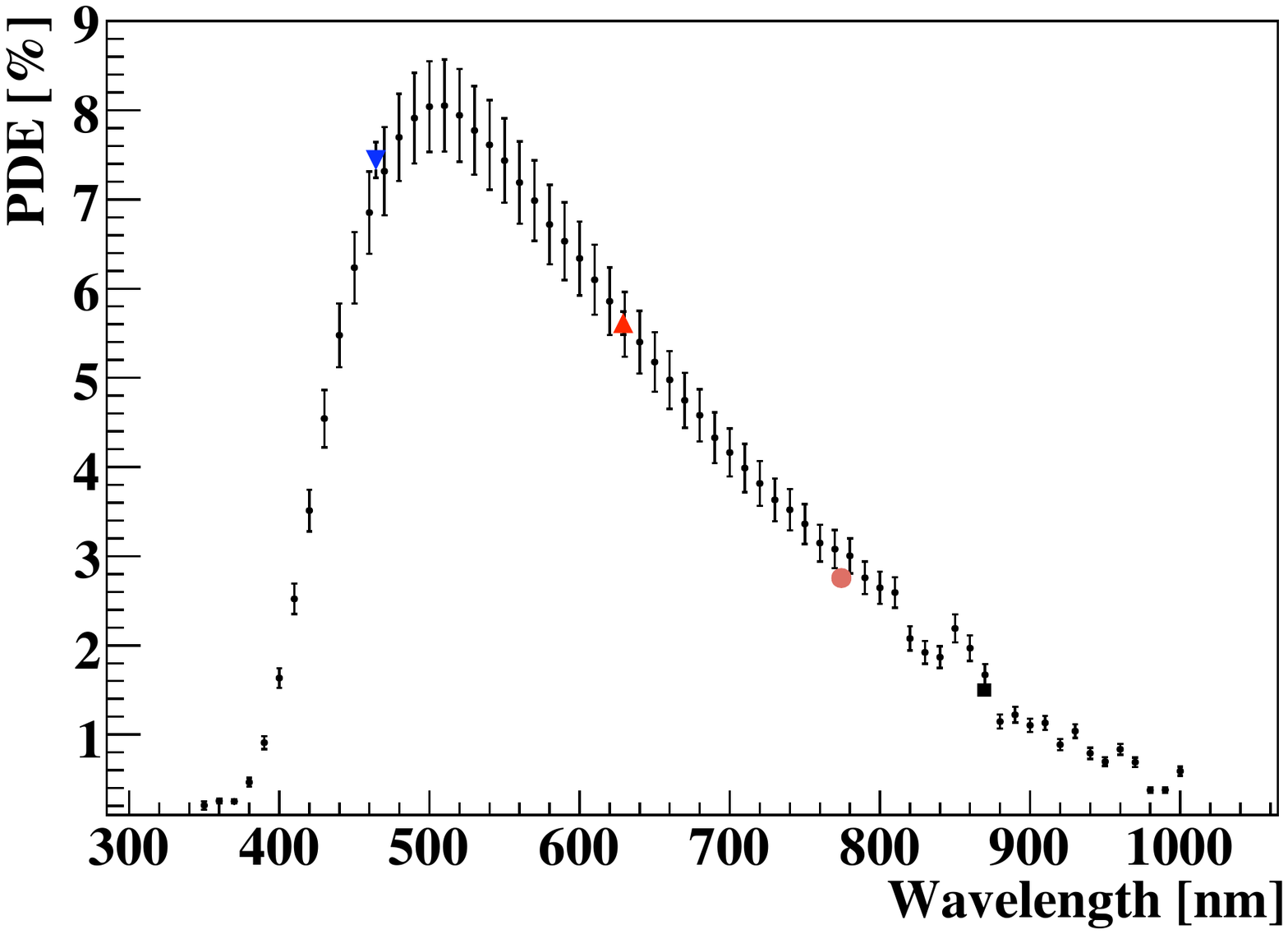}
	\caption{{\bf (left)} Photon detection efficiency of the SensL SPMMICRO1020X13 as a function of the over voltage for different wavelengths of light. {\bf (right)} PDE as a function of the wavelength at an over voltage of $\mathrm{U_{\rm{over}}=(2.5 \pm 0.05)\,V}$ at room temperature ($25\pm1.5 \, ^{\circ}\mathrm{C}$).}
	\label{fig:pde_sensl}
\end{figure*}

The PDE measurements for four different types of SiPM sensors are shown in Figure \ref{fig:pde1600} to \ref{fig:pde_sensl}. The diagrams on the left show the Photon Detection Efficiency as a function of the applied over voltage. The range of $U_{{\rm over}}$ was chosen such that PDE measurements are only presented for voltages for which a clear separation of the photon peaks in the QDC spectra (Fig.\ \ref{fig:histo}) is observed. This is essential for the statistical analysis applied for these measurements (cf. section \ref{sec:stat}).

A clear increase of the PDE with increasing over voltage is observed. This is mainly due the voltage dependence of the trigger efficiency, $\mathrm{\epsilon_{trigger}}$ , in equation \ref{eq:geiger}. Saturation of the PDE sets in once $\mathrm{\epsilon_{trigger}}$ approaches its maximum. A similar behaviour has been observed in \cite{uozumi}.

The Photon Detection Efficiencies as a function of wavelength are displayed on the right of Figures \ref{fig:pde1600} to \ref{fig:pde_sensl}. The diagrams show the relative sensitivity measurements normalised to the absolute PDE values at $633\,$nm determined with the statistical method using pulsed light from a laser diode. The $633\,$nm result is chosen for normalisation as the spectral width of the emission spectrum is small ($\mathrm{\sim 5 \, nm}$ FWMH) for the corresponding laser diode; for the LEDs this width lies between 10 and $20\,$nm. The absolute PDE measurements for 465, 775 and 870 nm are also shown; good agreement is observed with the normalised relative sensitivity curves.

The three SiPM sensors from HAMAMATSU show a large PDE in the blue spectral range peaking around $450\,$nm. This high blue sensitivity is due to their p-over-n structure. Because of its high absorption coefficient blue light cannot penetrate deeply into the silicon, and electron-hole pairs produced by it can only be generated close to the surface of the device. For p-over-n structures this is near the p-side of the depletion layer for which the probability of electrons to trigger an avalanche is largest, as has been shown in \cite{oldham}; due to the different ionisation coefficients electrons are more likely to trigger an avalanche than holes. Vice versa, in case of an n-over-p structure, as implemented for the SensL SPMs, most electron-hole pairs produced by blue light will be generated close to the n-side of the depletion layer resulting in a reduced efficiency in the blue thus explaining the maximum of the PDE in the green spectral range.

As reported previously \cite{bonanno,miyamoto,vacheret,korpar,musienko,dinu}, we also observe that the measured peak PDE values differ significantly from the reference values quoted by the manufacturer HAMAMATSU of 25\% (-025C), 50\% (-050C) and 65\% (-100C) \cite{hamamatsu}. The peak PDE values determined here (cf.\ Figures \ref{fig:pde1600}-\ref{fig:pde100}, right) are smaller than the reference values by 24\% (-025C), 36\% (-050C) and 43\% (-100C). Most of this discrepancy seems to be due to the measurement method used to derive the producer reference data which includes the contributions from optical cross-talk and after-pulses. As is shown in section \ref{sec:cross} and \ref{sec:afterp}, the cross-talk and after-pulse probabilities increase significantly for increased pixel sizes which explains why the discrepancy is larger for larger pixels.

The main systematic uncertainties of the presented measurements are given by the uncertainty on the power ratio (cf.\ Table \ref{table:R} and Figure\ \ref{fig:ratio}) and the accuracy of the calibrated reference photodiode as well as the ampere meter used in the relative sensitivity. For the absolute PDE determination the measurements are repeated several times to estimate the statistical uncertainty.

\section{Cross-talk Measurement}
\label{sec:cross}

Already in 1955 it was shown that a p-n junction which is reversely biased until breakdown emits light in the visible range \cite{newman}. Lacaita et al.\ later have shown that the efficiency for photon emission with energies larger than $1.14\, \mathrm{eV}$ is about $3 \cdot 10^{-5}$ per charge carrier crossing the junction \cite{lacaita}. Thus, assuming a typical SiPM gain of $10^6$, one finds that on average 30 such photons are generated during a pixel breakdown. Depending on their energy and the location where they are produced, these photons have a certain probability to reach a neighbouring pixel and to produce an additional avalanche.
The corresponding signal cannot be separated from a signal induced by an initial photon. This cross-talk limits the photon-counting resolution of SiPM devices as it is impossible to determine the exact number of photon-induced pixel-breakdowns. The cross-talk probability is thus an important characteristic of a SiPM and should be as small as possible. 

The method to determine the cross-talk is based on the analysis of signal events generated by thermal or field mediated excitations of electrons in the silicon lattice, often referred to as thermal noise or dark rate.

For an ideal detector the probability of two or more simultaneous thermal excitations should be negligible such that without cross-talk only signals with an amplitude corresponding to a single photoelectron ($\mathrm{1\,pe}$) should be observed. However, for a real SiPM additional, cross-talk induced avalanches can occur resulting in higher amplitude pulses. By comparing the event rate above a $\mathrm{1\,pe}$ threshold with the total dark rate measured the cross-talk probability is estimated.
\subsection{Experimental Setup}
\label{sec:csetup}
\begin{figure}[tbp]
	\centering
	\includegraphics[width=1\linewidth]{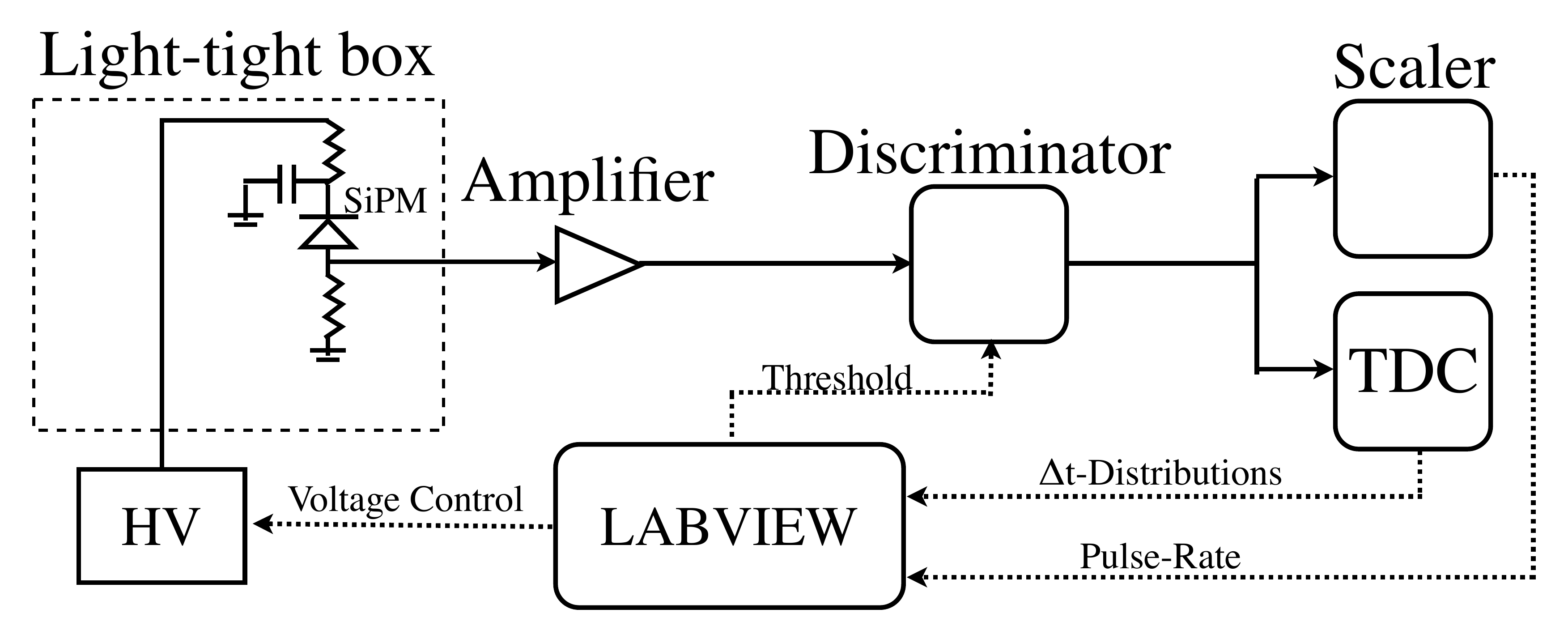}
	\caption{Experimental setup for cross-talk and after-pulse measurements. For the after-pulse measurement the discriminator output is connected to a TDC.}
	\label{fig:setcross}
\end{figure}

\begin{figure}[tbp]
	\centering
	\includegraphics[width=1\linewidth]{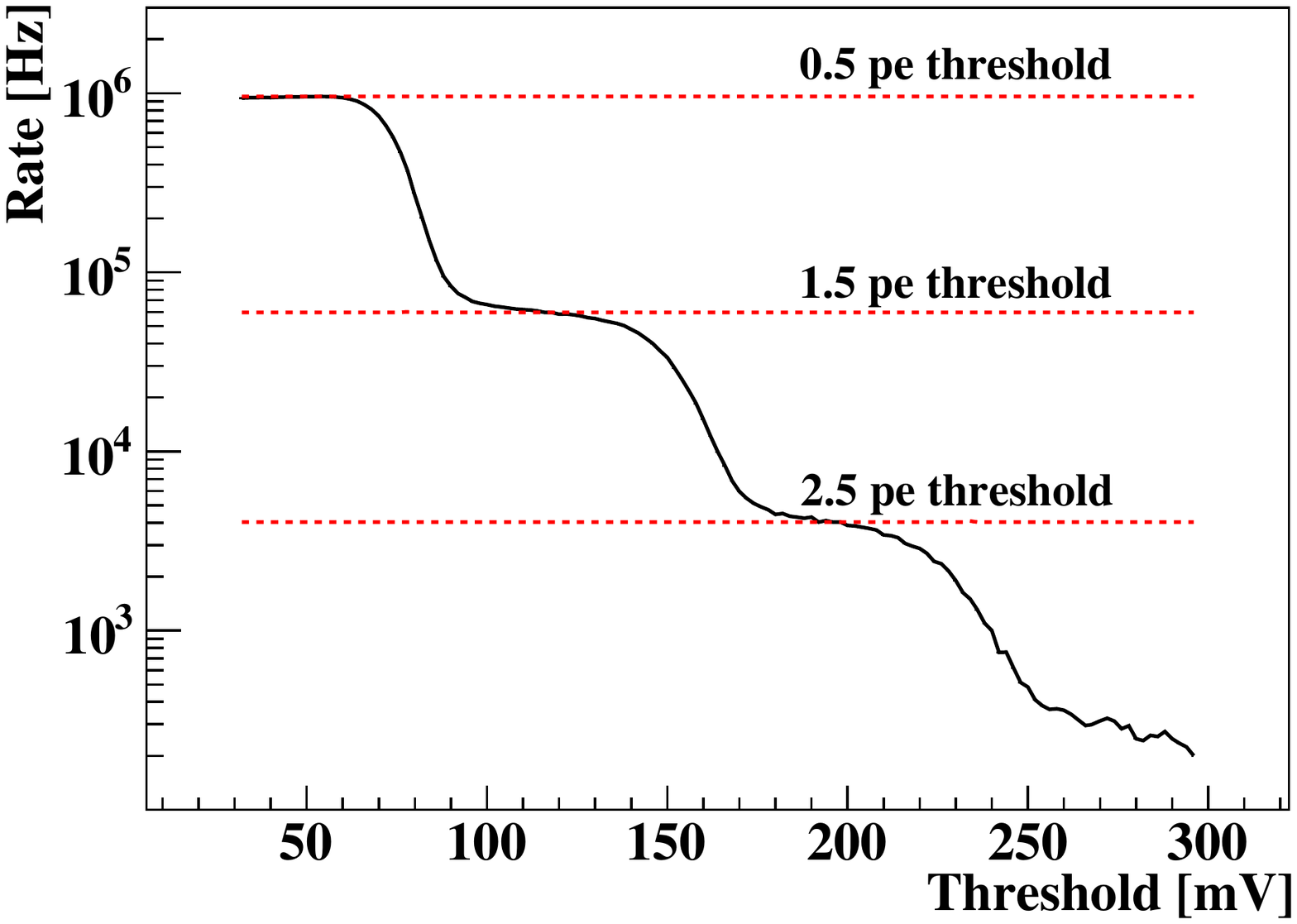}
	\caption{Thermal noise rate of a HAMAMATSU S10362-11-050C operated at $U_{{\rm over}}=1.3\,$V as a function of the discriminator threshold. The noise rates at the 0.5, 1.5 and 2.5 photoelectron threshold are indicated by the horizontal lines.}
	\label{fig:threshold}
\end{figure}

For the cross-talk measurement the SiPMs are placed in a light-tight box (cf.\ Figure  \ref{fig:setcross}) to prevent any background contamination from ambient light. The thermal noise signal is then amplified by a factor of 50 and fed into a discriminator module\footnote{LeCroy Model 4416}. Each time a noise pulse crosses a defined threshold level, a logical output pulse of $3\,$ns length is generated. These pulses are counted by a scaler module\footnote{LeCroy Model 2550B} with its time gate set to $1\,\mathrm{s}$. The measured count rate as a function of the discriminator threshold is shown in Figure \ref{fig:threshold}; in the presented example $U_{{\rm over}}$ was set to $1.3\,$V.

A characteristic step function is observed, and the count rate drops every time integer multiples of the $\mathrm{1\,pe}$ threshold are reached. For a threshold above the electronic noise but below the amplitude of the $\mathrm{1\,pe}$ signal all SiPM thermal or field mediated excitations are counted; this corresponds to the dark rate, $\mathrm{\nu_{0.5\,pe}}$. By measuring the count rate above a $\mathrm{1.5\,pe}$ threshold, $\mathrm{\nu_{1.5 \,pe}}$, only events with one or more additional, cross-talk induced avalanche are taken into account. The ratio $\mathrm{P_{c}=\nu_{1.5\,pe} / \nu_{0.5 \,pe}}$ measures the cross-talk probability.

To determine the cross-talk probability as a function of the over voltage, several threshold spectra are recorded at different $U_{{\rm over}}$ settings. The rates $\mathrm{\nu_{0.5\,pe}}$ and $\mathrm{\nu_{1.5\,pe}}$ are determined automatically by fitting a spline to the data and calculating the absolute value of its derivative; this corresponds to the pulse-hight spectrum of thermal noise events. The first local minima of these spectra indicate the $\mathrm{0.5\,pe}$ and $\mathrm{1.5\,pe}$ threshold values at which the count rates $\mathrm{\nu_{0.5\,pe}}$ and $\mathrm{\nu_{1.5\,pe}}$ are determined. The measurement uncertainty is roughly estimated by varying the threshold by $50\, \%$ of the the plateau width.

 \begin{figure}[htbp]
	\centering
	\includegraphics[width=1\linewidth]{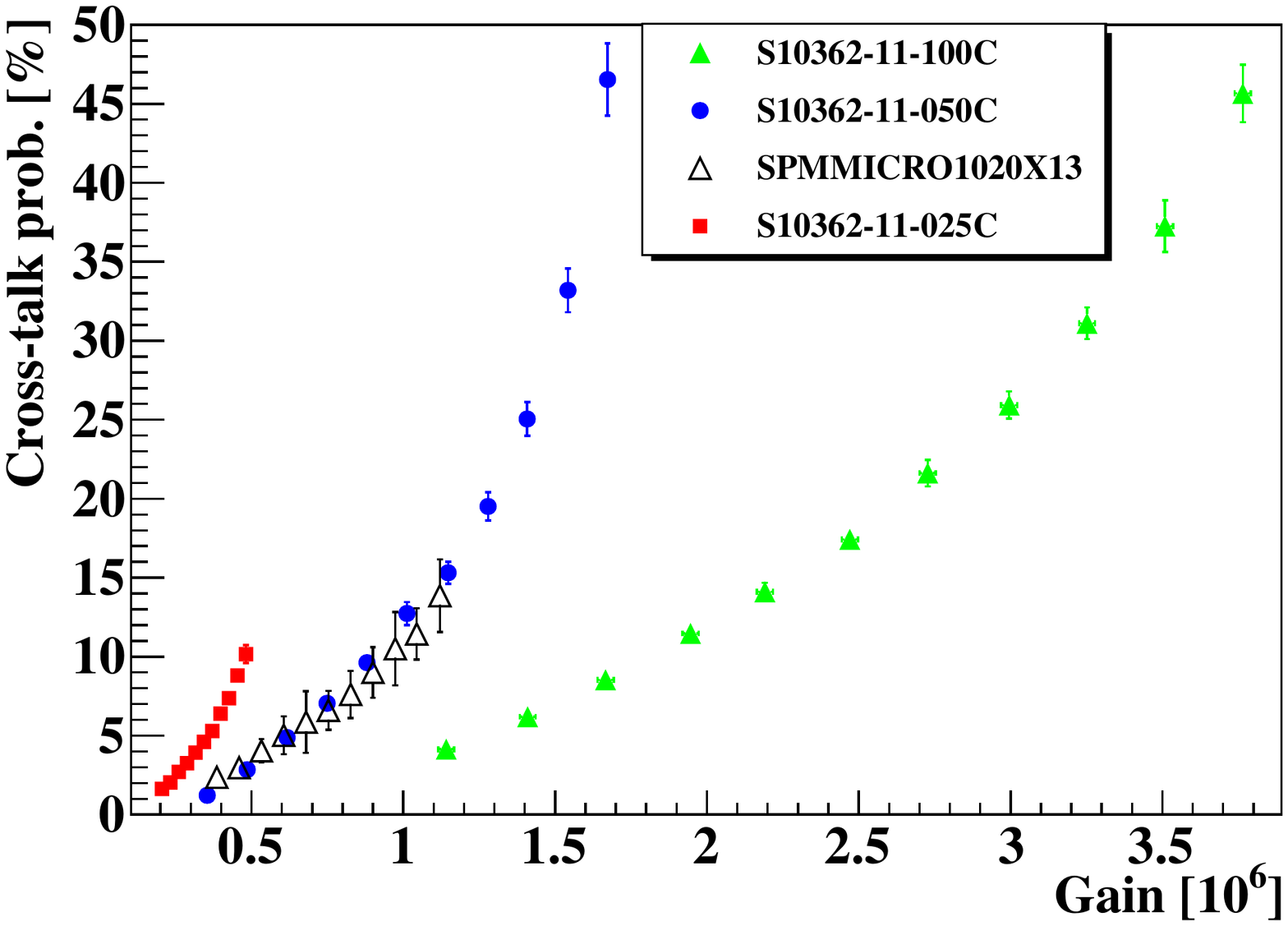}
	\caption{Cross-talk probability for different SiPM sensors as a function of the SiPM gain.}
	\label{fig:crosstalk}
\end{figure}
\subsection{Cross-talk Results}
The measured cross-talk probability as a function of the SiPM gain is shown in Figure  \ref{fig:crosstalk} for the tested SiPM devices; as expected, it increases with increasing gain and thus increasing over voltage $U_{{\rm over}}$ \mbox{(cf.\ Figure \ref{fig:gain})}. If more charge carriers traverse the pn-junction, more photons are produced during the breakdown process yielding a higher cross-talk probability; this effect is additionally enhanced by a larger avalanche trigger efficiency, $\mathrm{\epsilon_{trigger}}$ which also rises with $U_{{\rm over}}$.

Comparing the cross-talk probability of the three MPPCs with different cell sizes at constant gain values, shows that devices with larger cells have a smaller cross-talk probability compared to devices with smaller cells. This can be explained by the different values of $\mathrm{\epsilon_{trigger}}$ at constant gain and by the longer average distance photons have to travel in case of larger cells before reaching a neighbouring pixel where they can cause a second avalanche.

The tested SensL device shows a different behaviour which is presumably caused by a different production technique. It has a similar cross-talk probability as the 400 pixel MPPC device (at the same gain) although featuring a higher pixel density (smaller cell size).

\section{After-Pulse Measurement}
\label{sec:afterp}
The measurement of the after-pulse probability is also based on the analysis of the noise rate as this not only includes thermal excitations but also after-pulsing. After-pulses are believed to be generated if electrons produced in an avalanche are trapped and released again after some delay which can last from nanoseconds up to several microseconds.
The charge fraction carried by these pulses depends on the recovery state of the corresponding pixel and can be calculated if the pixel recovery time $\tau_{{\rm r}}$ is known: $\xi(\Delta {\rm t})=1-exp(-\Delta {\rm t / \tau_r})$.
If the time delay with respect to the preceding pulse, $\Delta {\rm t}$, is short, only pulses with small, i.e.\ smaller than the $1\,$pe signal amplitude are generated; if the delay is larger than the pixel recovery time, a standard avalanche signal is triggered. These signals cannot be separated from genuine, photon-induced signals and thus deteriorate the photon-counting resolution. 

 \begin{figure}[tbp]
	\centering
	\includegraphics[width=1\linewidth]{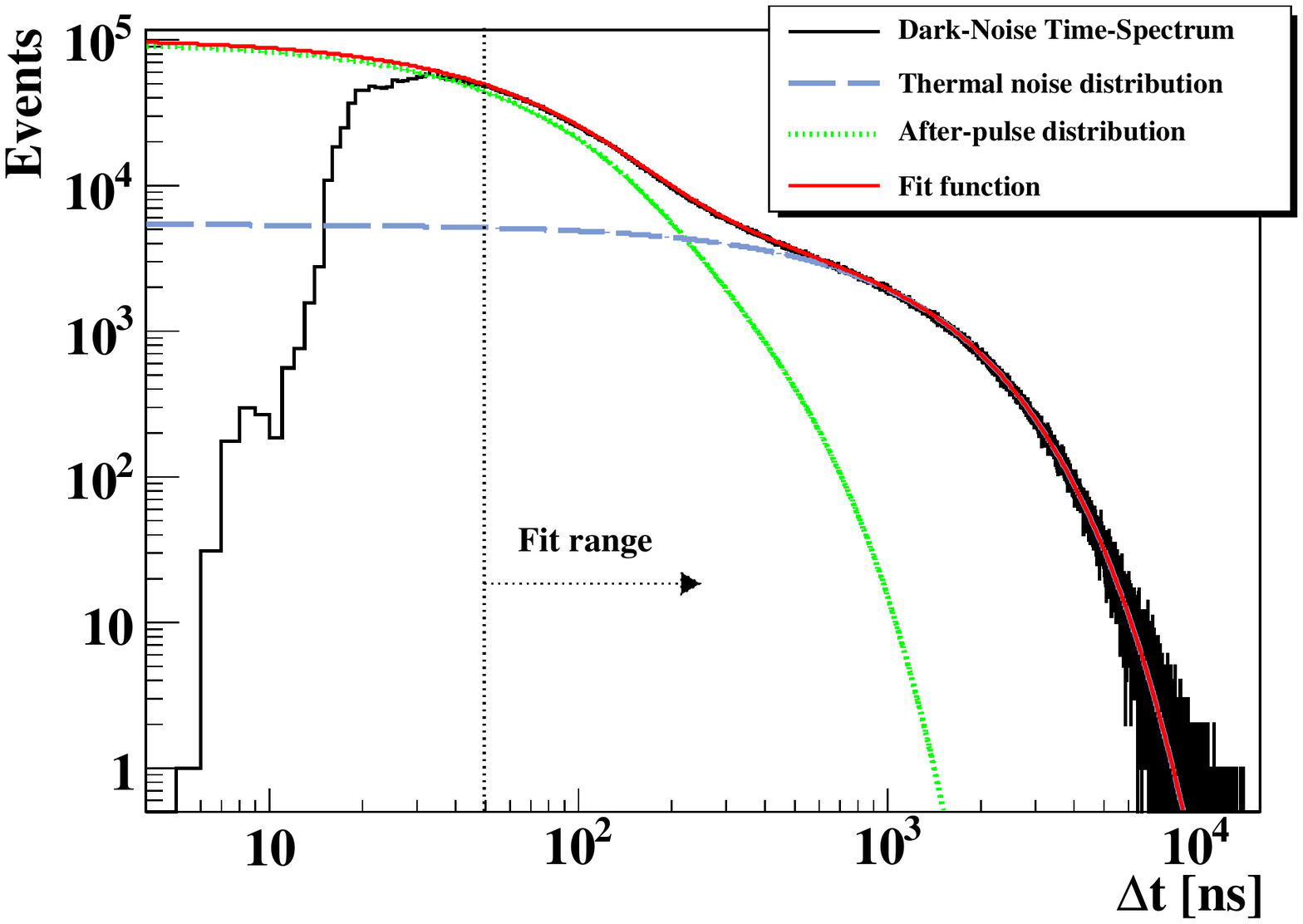}
	\caption{After-pulse time difference distribution of a HAMAMATSU S10362-11-050C at a bias voltage of $-70.6\,$V. The distribution is well represented by a fit-function (red line) which is given by the superposition of two exponential one for the thermal noise and one for the after-pulse time distribution.}
	\label{fig:afterpulse_dist}
\end{figure}

\subsection{Experimental Setup}
The experimental setup is similar to the one used for the cross-talk determination (cf.\ Figure \ref{fig:setcross}). The only difference is that the discriminator signals are now fed into a TDC\footnote{CAEN V1290A, Multihit TDC} instead of a scaler module. The TDC is used to measure the time difference, $\mathrm{\Delta t}$, between consecutive SiPM pulses. A typical $\mathrm{\Delta t}$-distribution is shown in Figure \ref{fig:afterpulse_dist}. For small values of $\mathrm{\Delta t}$ the efficiency for detecting an after-pulse is largely reduced due to the time needed for pixel recovery and the dead time of discriminator and TDC. For time differences with $\xi (\Delta {\rm t}) \approx 1$, i.e.\ larger than $20-100\,$ns - depending on the sensor type and the applied over voltage - the measured distribution can be fitted by a superposition of two exponentials:

\begin{align}
\mathrm{n_{tp}(\Delta t)}&=\mathrm{N_{tp} / \tau_{tp} \cdot e^{-\frac{\Delta t}{\tau_{tp}}}} \label{eq:tp}\\
\mathrm{n_{ap}(\Delta t)}&=\mathrm{ N_{apf} / \tau_{apf} \cdot e^{-\frac{\Delta t}{\tau_{apf}}} + N_{aps} / \tau_{aps} \cdot e^\frac{-\Delta t}{\tau_{aps}}}.
\label{eq:ap}
\end{align}

Here, equation \ref{eq:tp} describes the probability density for thermal events, with the constant $\mathrm{N_{tp}}$ corresponding to the integrated number of thermal signals and $\mathrm{1/ \tau_{tp}}$ representing the reduced dark count rate (without after-pulses).
The probability density for after-pulses is given by equation \ref{eq:ap}. As already observed in \cite{retiere_1}, the fit quality can be significantly improved by using two different time constants, $\mathrm{\tau_{apf}}$ and $\mathrm{\tau_{aps}}$, one describing a fast component of after-pulse generation and the other a slow one. $\mathrm{N_{apf}}$ and $\mathrm{N_{aps}}$ correspond to the integrated number of fast and slow after-pulses, respectively. The after-pulse probability is then given by:

\begin{equation}
\mathrm{P_{ap} = \frac{ \int_{0}^\infty \xi \cdot n_{ap}\, d\Delta t} { \int_{0}^\infty \xi \cdot(n_{ap}+n_{tp})\, d\Delta t }},
\end{equation}

where $\xi$, $\mathrm{n_{ap}}$ and $\mathrm{n_{tp}}$ depend on $\mathrm{\Delta t}$. The function $\xi$ takes into account that trapped electrons which are released prior to complete pixel recovery have a smaller contribution to the after-pulse probability. Since the left part of the spectrum (cf.\ Fig.\ \ref{fig:afterpulse_dist}) was not used in the fit, the recovery time\footnote{100 pix. $\mathrm{ \tau_r=33\,}$ns, 400 pix. $\mathrm{ \tau_r=9\,}$ns, \mbox{1600 pix. $\mathrm{ \tau_r=4\,}$ns}} $\mathrm{\tau_r}$ was taken from \cite{oide}.
\subsection{Results}
\begin{figure}[tbp]
	\centering
	\includegraphics[width=1\linewidth]{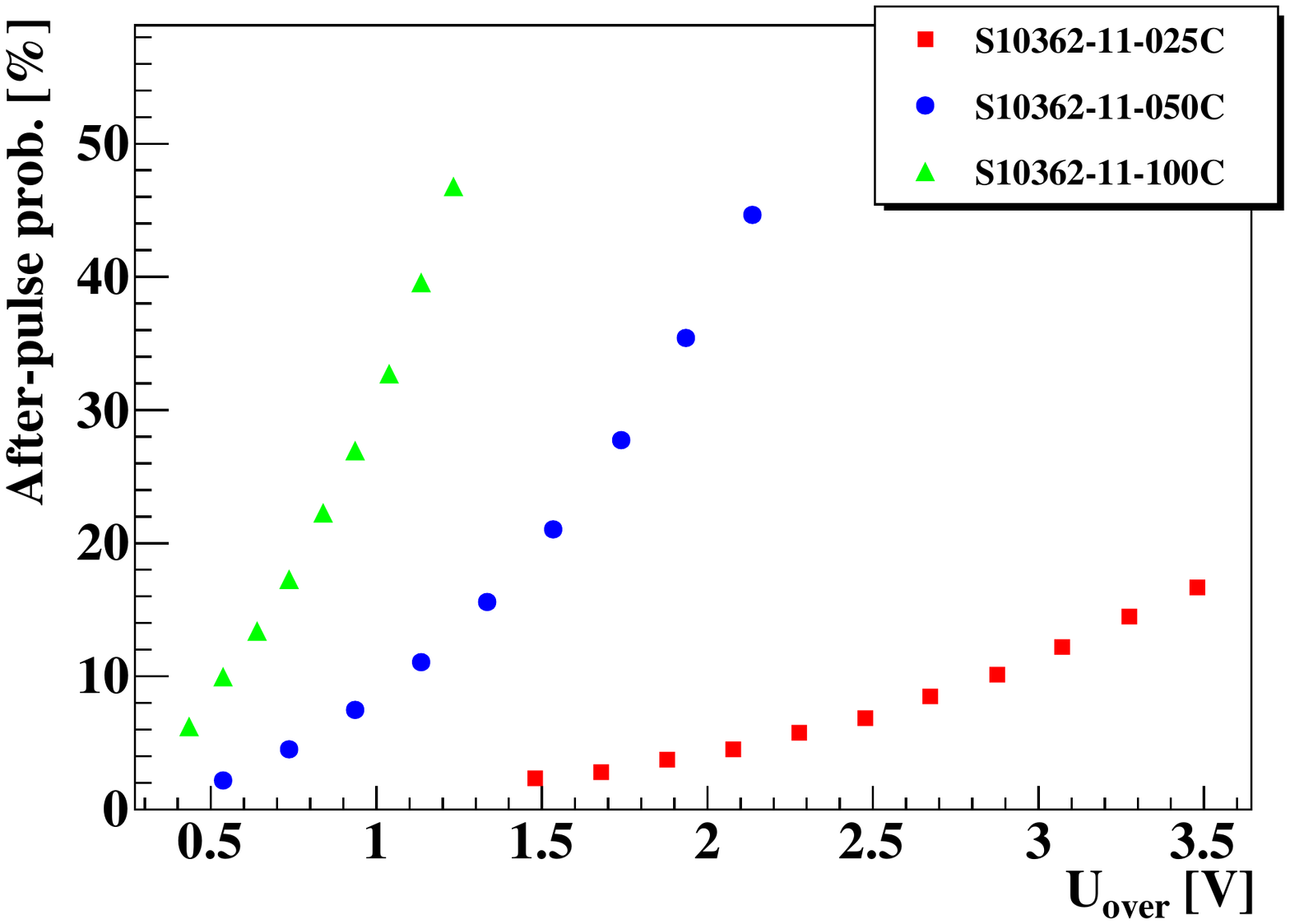}
	\caption{After-pulse probability as a function of the over voltage measured for three different sensor types.}
	\label{fig:afterpulse}
\end{figure}

The after-pulse probability as a function of the over voltage is shown in Figure \ref{fig:afterpulse}; it increases with increasing $U_{{\rm over}}$. The reason for this increase is again due to the increase in gain. The second effect which gives rise to the super-linear increase is as before caused by a rise of the avalanche trigger probability.

For the tested SensL SPM, the measured dark-noise time spectrum is well described by a thermal noise contribution only (equation \ref{eq:tp}). Hence, the determined after-pulse probability is negligible small. This is induced by the relatively long pixel recovery time of the device causing that trapped charge carriers which are successively released do not generate after-pulses.  

\section{Uniformity Scans}
\label{sec_uni}
By raster scanning the active area of SiPMs with a small laser spot the spatial uniformity of the devices is studied. A similar measurement focussing on the spatial variance of the photon sensitivity for different SiPM types was performed in \cite{krizan}. The method presented here also allows to determine the uniformity in gain as well as a position dependent cross-talk probability.

\begin{figure}[tbp]
	\centering
	\includegraphics[width=1\linewidth]{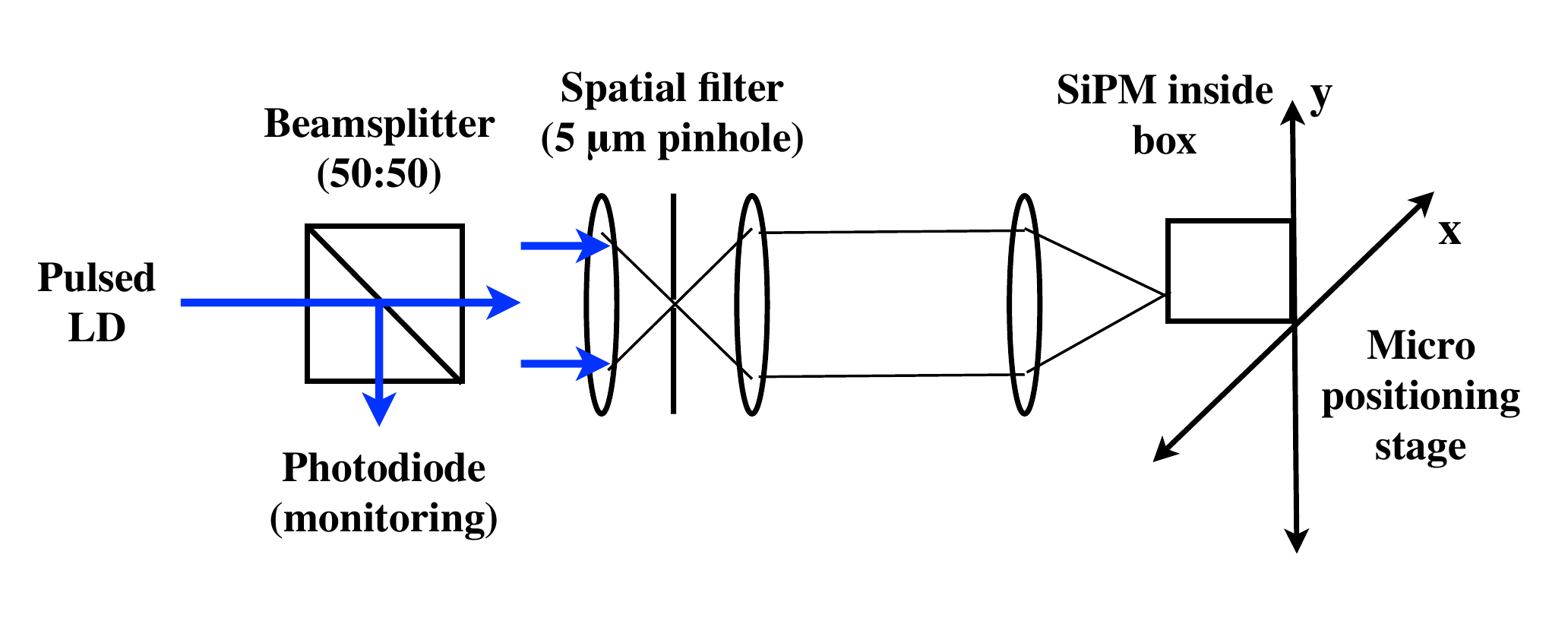}
	\caption{Schematic view of the experimental setup for raster scanning SiPMs.}
	\label{fig:uniformity_setup}
\end{figure}

The experimental setup used for the uniformity studies is schematically depicted in Figure \ref{fig:uniformity_setup}. A laser diode generates short light pulses of about $2\,$ns length which are split by a beam splitter into two separate beams. One of the beams is monitored by a photodiode for long term intensity variations while the other passes a spatial filter and is focussed onto the active area of a SiPM.

\subsection{Experimental Setup}

\begin{figure}[tbp]
	\centering
	\includegraphics[width=1\linewidth]{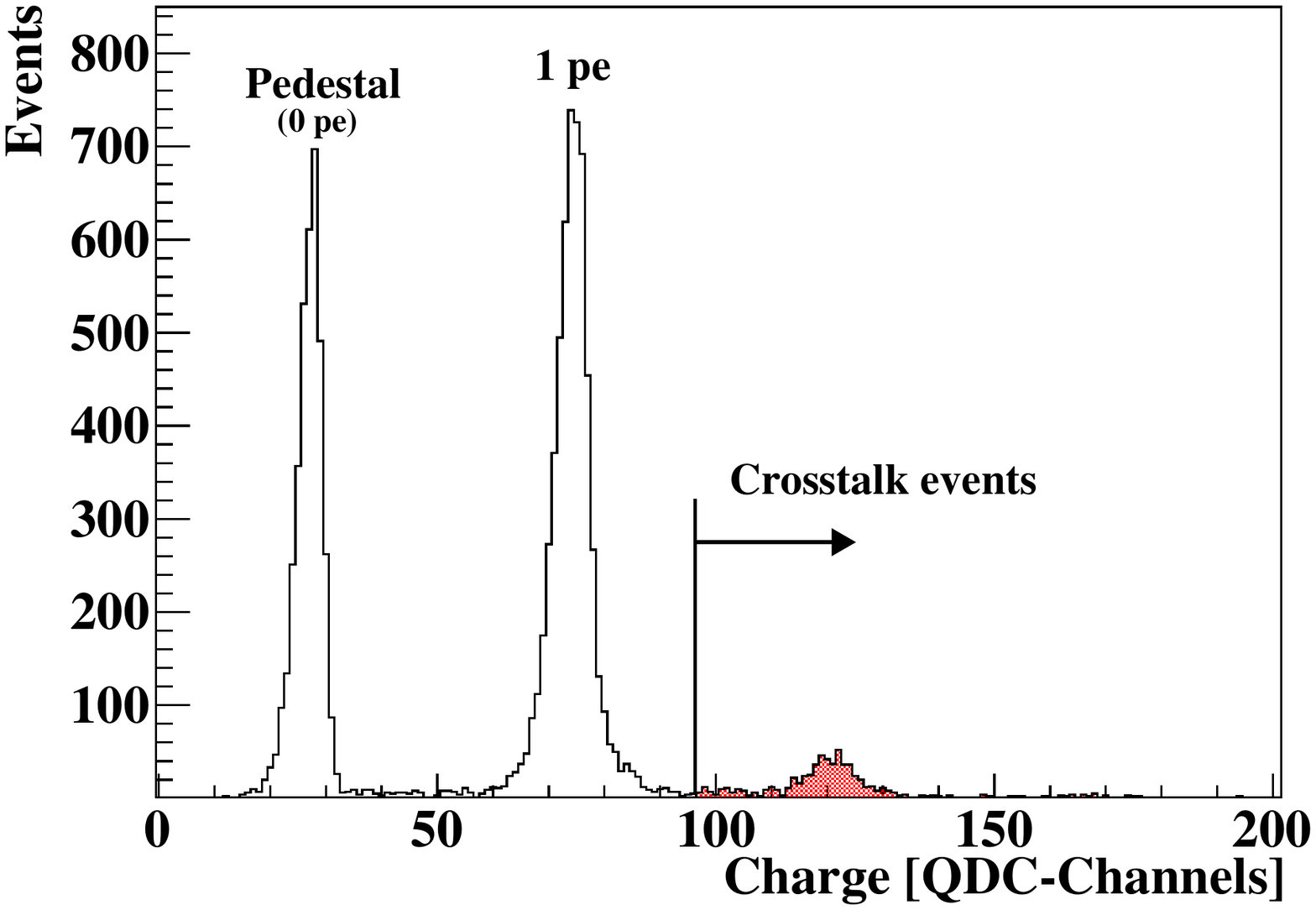}
	\caption{Sample charge histogram recorded for the uniformity scan measurement. The shaded area indicates the number of cross-talk events.}
	\label{fig:spectrum_array}
\end{figure}

The SiPM is installed inside an aluminium box mounted on a $xy$-positioning stage such that it can be scanned by the light spot with $1\, \mu$m precision. The box shields the SiPM from electrical noise mainly caused by the operation of the positioning stage. At each geometrical position a charge histogram (Fig.\ \ref{fig:spectrum_array}) with 10,000 entries is recorded and evaluated using the same analysis method described in section \ref{absolute}.

The setup allows to measure the position dependent sensitivity without any influence from cross-talk and after-pulsing. Furthermore, gain and cross-talk probability can be calculated for individual pixels using the recorded charge histograms. The latter is defined here as the ratio of events found above the single photoelectron peak, $N_{{\rm c}}$ (shaded area in Fig.\ \ref{fig:spectrum_array}) to the total number of signal events above the pedestal:

\begin{equation}
\mathrm{P^{px}_{c}}=\frac{N_{{\rm c}}-N_{{\rm 1pe}}\cdot (N^{{\rm dark}}_{{\rm tot}}/N^{{\rm dark}}_{{\rm ped}}-1)}{N_{{\rm c}}+N_{{\rm 1pe}}}.
\label{eq:pxct}
\end{equation}

The second term in the numerator is a correction which accounts for the cross-talk events caused by thermal noise (cf.\ equation \ref{eq_stat1} and Fig.\ \ref{fig:darkrate}), and $N_{{\rm 1pe}}$ represents the number of events in the $1\,$pe peak. In the absence of optical cross-talk, histogram entries exceeding $1\,$pe (cf. Fig. \ref{fig:spectrum_array}) would not be expected using the described experimental setup, as only single pixels are illuminated and the light pulse width is tuned to be much smaller than the pixel recovery time. The fact that also $2\,$pe, $3\,$pe etc.\ events are observed indicates the existence of optical cross-talk. For the measurement a short QDC gate of about $30\,$ns was chosen in order to keep the influence of fast after-pulses small.

\subsection{Uniformity Scan Results}
\begin{figure*}[htbp]
	\centering
	\includegraphics[width=0.32\linewidth]{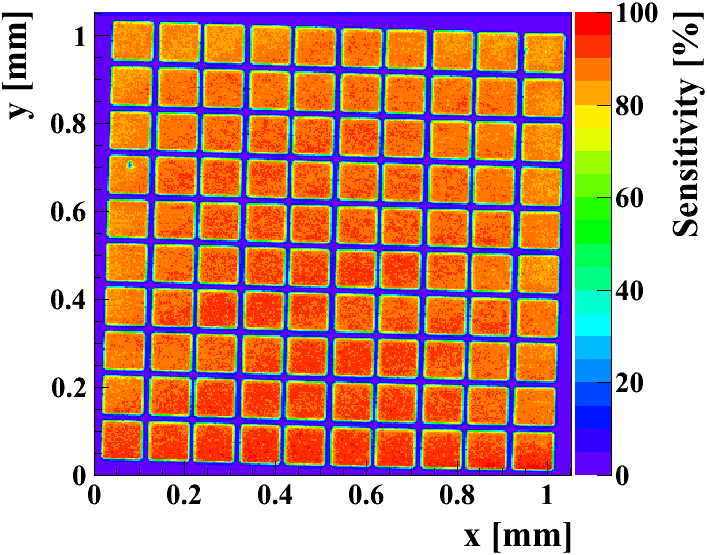}
	\hspace{0.1cm}
	\includegraphics[width=0.32\linewidth]{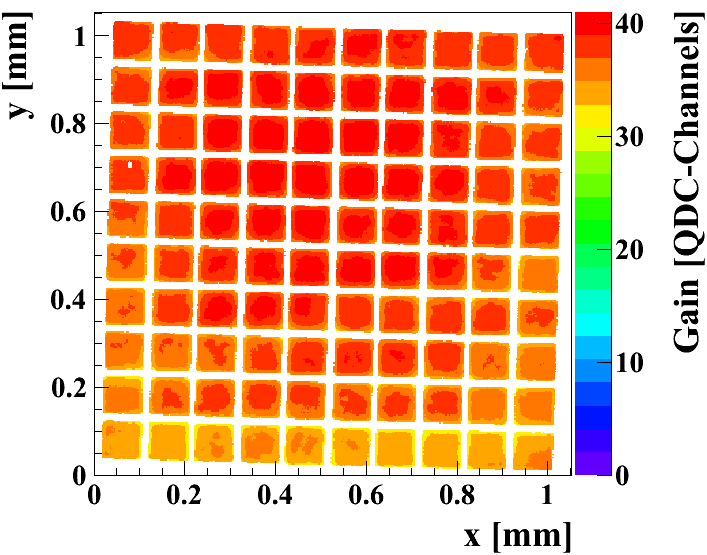}
	\hspace{0.1cm}
	\includegraphics[width=0.32\linewidth]{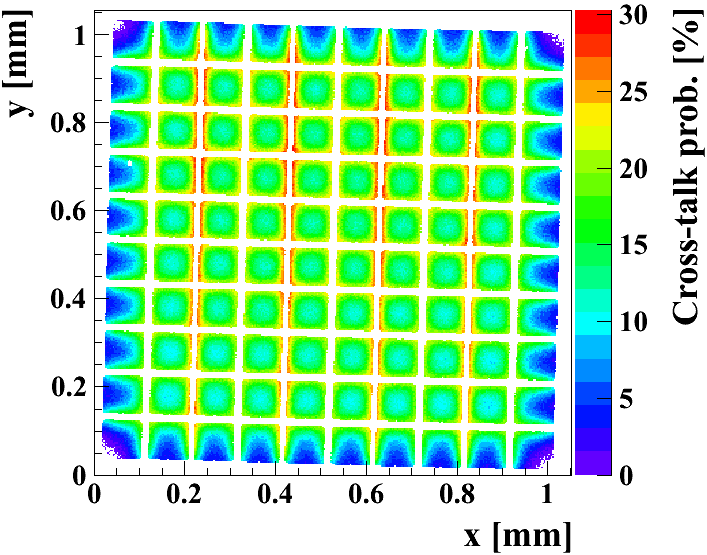}
	\caption{Results of the uniformity scan measurements for a HAMAMATSU S10362-11-100C operated at $\mathrm{U_{\rm{over}}=1.1\,V}$: {\bf (left)} relative sensitivity map, \mbox{{\bf (centre)} gain map}, {\bf (right)} cross-talk map. In case of the gain and cross-talk maps, geometrical positions with less than $50\%$ sensitivity are set to zero.}
	\label{fig:map_100}
\end{figure*}

\begin{figure*}[htbp]
	\centering	
	\includegraphics[width=0.32\linewidth]{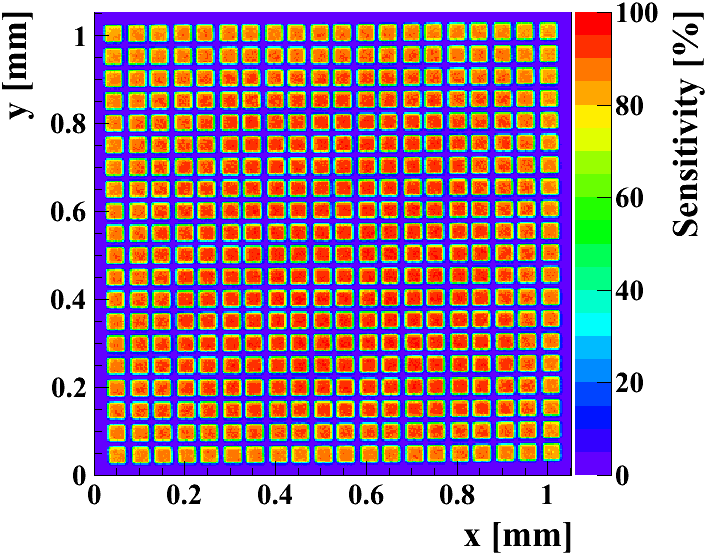}
	\hspace{0.1cm}	
	\includegraphics[width=0.32\linewidth]{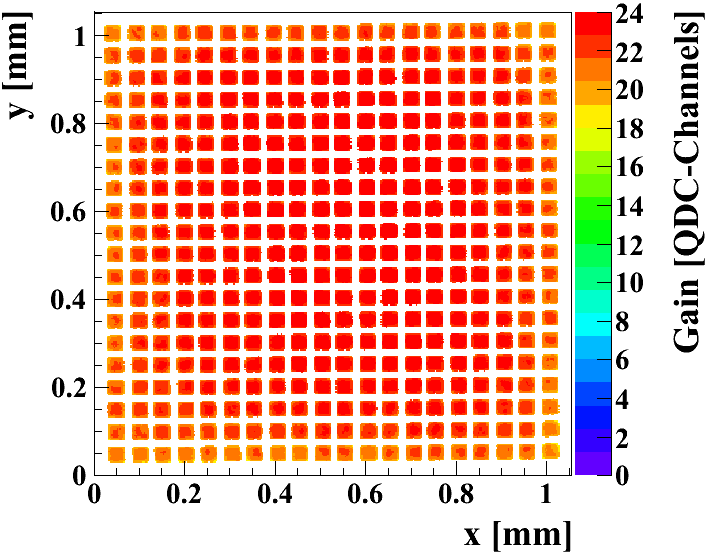}
	\hspace{0.1cm}	
	\includegraphics[width=0.32\linewidth]{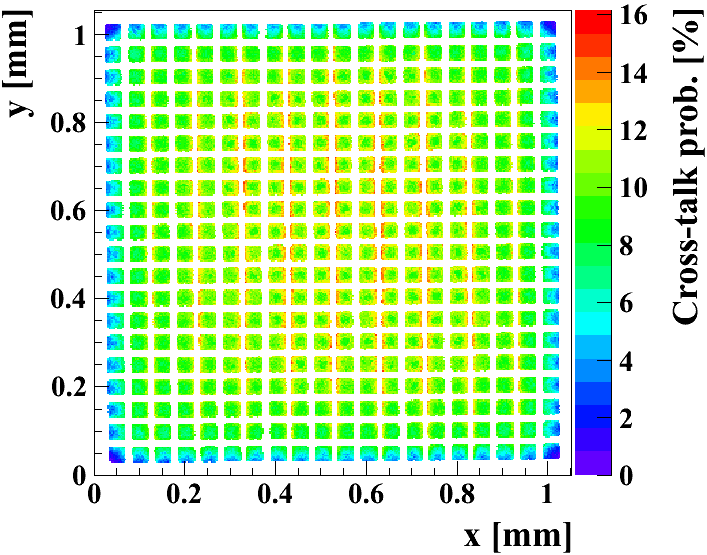}
	\caption{Results of the uniformity scan measurements for a HAMAMATSU S10362-11-050C operated at $\mathrm{U_{\rm{over}}=1.3\,V}$: {\bf (left)} relative sensitivity map, \mbox{{\bf (centre)} gain map}, {\bf (right)} cross-talk map. In case of the gain and cross-talk maps, geometrical positions with less than $50\%$ sensitivity are set to zero.}
	\label{fig:map_400}
\end{figure*}

\begin{figure*}[htbp]
	\centering	
	\includegraphics[width=0.32\linewidth]{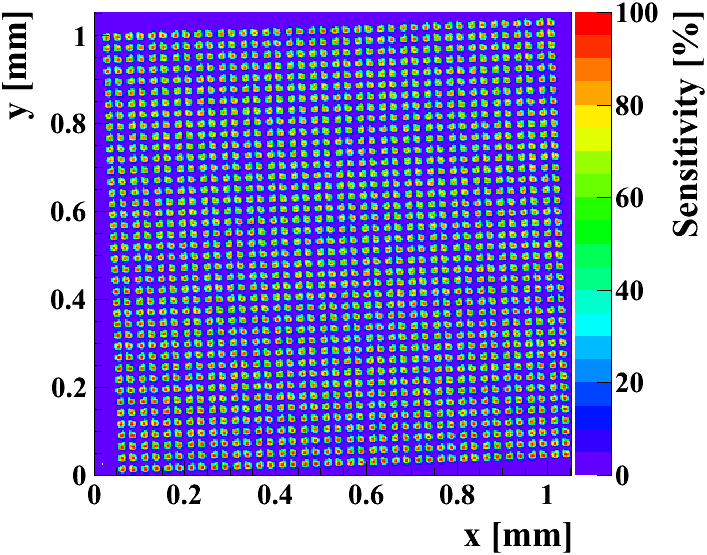}
	\hspace{0.1cm}	
	\includegraphics[width=0.32\linewidth]{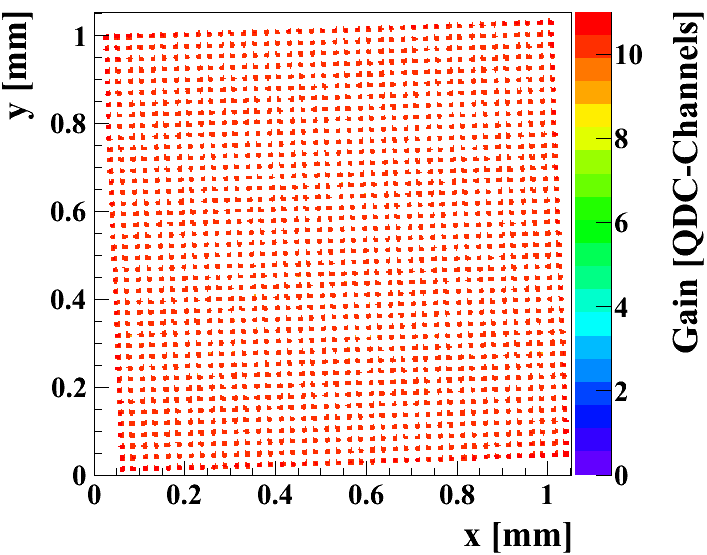}
	\hspace{0.1cm}	
	\includegraphics[width=0.32\linewidth]{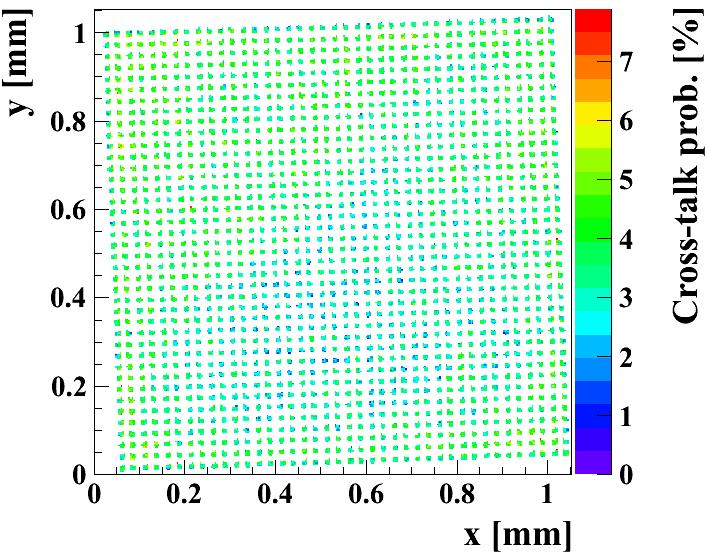}
	\caption{Results of the uniformity scan measurements for a HAMAMATSU S10362-11-025C operated at $\mathrm{U_{\rm{over}}=2.3\,V}$: {\bf (left)} relative sensitivity map, \mbox{{\bf (centre)} gain map}, {\bf (right)} cross-talk map. In case of the gain and cross-talk maps, geometrical positions with less than $50\%$ sensitivity are set to zero.}
	\label{fig:map_1600}
\end{figure*}

\begin{figure*}[htbp]
	\centering	
	\includegraphics[width=0.32\linewidth]{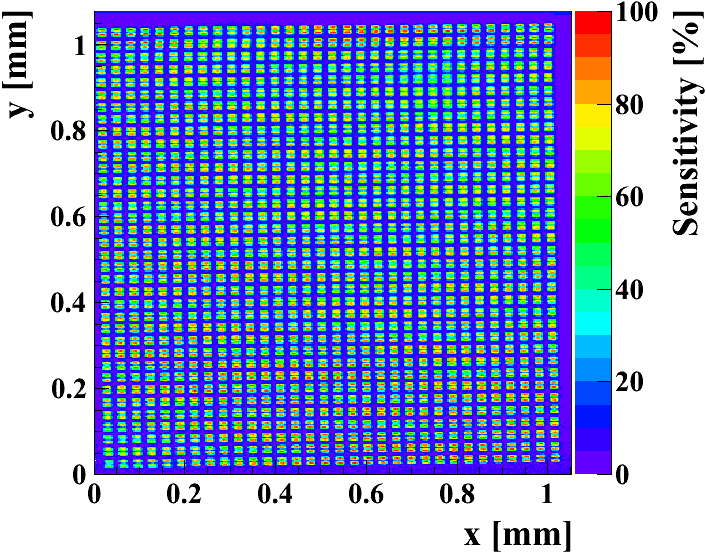}
	\hspace{0.1cm}	
	\includegraphics[width=0.32\linewidth]{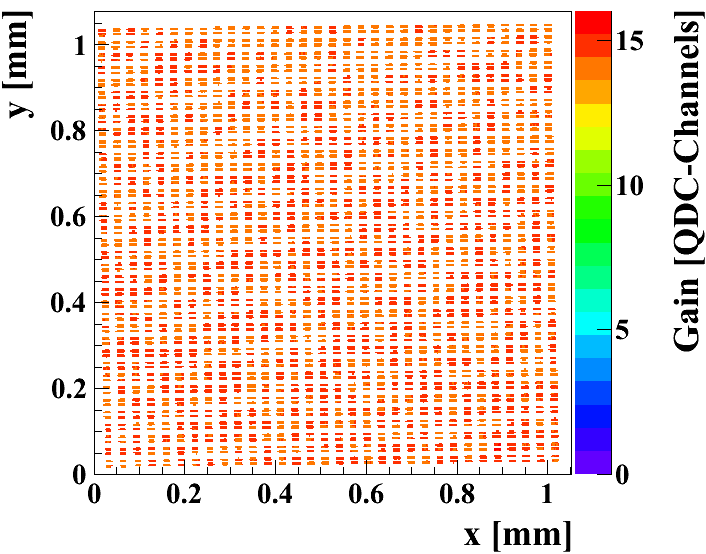}
	\hspace{0.1cm}	
	\includegraphics[width=0.32\linewidth]{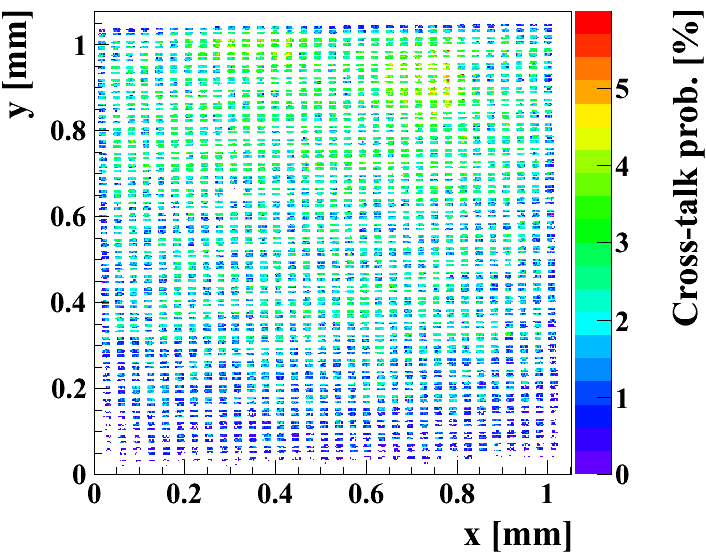}
	\caption{Results of the uniformity scan measurements for a Sensl SPMMICRO1020X13 operated at $\mathrm{U_{\rm{over}}=1.6\,V}$: {\bf (left)} relative sensitivity map, \mbox{{\bf (centre)} gain map}, {\bf (right)} cross-talk map. In case of the gain and cross-talk maps, geometrical positions with less than $50\%$ sensitivity are set to zero.}
	\label{fig:map_sensl}
\end{figure*}

The results of the uniformity scan measurement for the tested SiPM devices are shown in Figures \ref{fig:map_100}-\ref{fig:map_sensl}; they show a high degree of homogeneity. Variations of about $10\, \%$ to $20\, \%$ are observed in gain and sensitivity.

For the cross-talk probability a clear dependence on the geometrical position is observed; it is small for pixels at the edge of the active area and gets substantially larger when moving into the centre of the device. This is due to the different number of neighbouring pixels: the more neighbours the higher the probability to induce an additional avalanche breakdown in one of them. For the devices with 50 and $100\, \mu$m pitch an additional variation on the single pixel scale is observed.

For the gain and single pixel cross-talk measurement a cut on $50\, \%$ sensitivity has been applied in order two guarantee a well-defined position determination. For low sensitivities - as it is given between pixels - the actual position of the firing primary pixel is unknown, which would prevent a reliable measurement.


\section{Conclusion and Discussion}
\label{conclusion}
An experimental test environment has been designed to determine the characteristics of Silicon Photomultipliers. The SiPM properties addressed here are the cross-talk and after-pulse corrected Photon Detection Efficiency in the spectral range between 350 and $1000\,$nm, the probability of cross-talk and after-pulse occurrence, and the spatial uniformity for sensitivity, gain and cross-talk. The presented measurement setups are and will be used to compare these properties for various SiPMs on the market in order to identify suitable sensors for applications in high-energy physics calorimetry, medical imaging and elsewhere.

Characteristics of several SiPM sensors from \mbox{HAMAMATSU} and SensL with maximum sensitivity in the blue and the green spectral region are presented and compared. The measured wavelengths of maximal PDE are consistent with those quoted by the producer. In case of the HAMAMATSU MPPCs significantly smaller PDE values than quoted by the producer have been measured. As the producers values are provided with the reference of including cross-talk and after-pulses, a difference is expected.

Cross-talk and after-pulse probabilities increase with the applied bias voltage and pixel capacitance as expected. In addition, all devices tested show a good spatial uniformity of sensitivity and gain. The position dependent cross-talk probability was found to be higher in the centre of a device than at its edges, compatible with the number of neighbouring pixels.

\section{Acknowledgements}
The authors wish to thank the CALICE collaboration, in particular Nicola D'Ascenzo, Erika Garutti, \mbox{Martin} G\"ottlich, Alexander Kaplan and Felix Sefkow for the good cooperation and the inspiring discussions.


\begin{thebibliography}{00}
\bibitem{andreev}
V. Andreev et al., Nucl. Instr. and Meth. A 564 (2006) 144-154.
\bibitem{astro}
A. Biland et al., Nucl. Inst. and Meth. A 595 (2008) 165-168.
\bibitem{pet}
S. Moehrs et al., Phys. Med. Biol. 51 (2006) 1113-1127.
\bibitem{renker}
D. Renker, E. Lorenz, JINST 4 P04004 (2009).
\bibitem{bonanno}
G. Bonanno et al., Nucl. Instr. and Meth. A 610 (2009) 93-97.
\bibitem{otte}
A. N. Otte et al., Nucl. Instr. and Meth. A 567 (2006) 360-363.
\bibitem{uozumi}
S. Uozumi et al., Nucl. Instr. and Meth. A 581 (2007) 427-432.
\bibitem{oldham}
W. G. Oldham et al., IEEE Trans. Electron Dev. 19 (1972) 1056-1060.
\bibitem{miyamoto}
H. Miyamoto et al., talk given at 11th Pisa meeting on advanced detectors, Elba, May 2009.
\bibitem{vacheret}
A. Vacheret et al., talk given at TIPP09, Tsukuba, March 2009.
\bibitem{korpar}
S. Korpar et al., Nucl. Instr. and Meth. A 613 (2010) 195-199.
\bibitem{musienko}
Y. Musienko et al., Nucl. Instr. and Meth. A 581 (2007) 433-437.
\bibitem{dinu}
N. Dinu et al., Nucl. Instr. and Meth. A 610 (2009) 423-426.
\bibitem{hamamatsu}
http://sales.hamamatsu.com/en/products/solid-state-division/si-photodiode-series/mppc.php, February 2010.
\bibitem{newman}
R. Newman et al., Phys. Rev. 100 (1955) 700 - 703.
\bibitem{lacaita}
A. Lacaita et al., IEEE Trans. Electron Dev. 40 (1993) 577-582.
\bibitem{retiere_1}
Y. Du, F. Reti\`ere, Nucl. Instr. and Meth. A 596 (2008) 396-401.
\bibitem{oide}
H. Oide et al., PoS(PD07)008 (2007).
\bibitem{krizan}
R. Pestotnik et al., Nucl. Instr. and Meth. A 581 (2007) 457-560.
\end{thebibliography}
\end{document}